\documentclass{emulateapj}
\usepackage{graphicx}
\usepackage{lscape}
\usepackage{epsfig}
\submitted{Accepted for Publication in AJ}
\shortauthors{West et al.}
\shorttitle{HI Selected Galaxies in the SDSS I}
\begin{document}

\title{HI Selected Galaxies in the Sloan Digital Sky Survey I: Optical Data}

\author{Andrew A. West\altaffilmark{1,2,3,4}, 
Diego A. Garcia-Appadoo\altaffilmark{5,6},
Julianne J.Dalcanton\altaffilmark{2}, 
Mike J. Disney\altaffilmark{6},
Constance M. Rockosi\altaffilmark{7},
{\v Z}eljko Ivezi{\'c}\altaffilmark{2},
Misty C. Bentz\altaffilmark{2,8},
J. Brinkmann\altaffilmark{9}}

\altaffiltext{1}{Corresponding author: aawest@bu.edu}
\altaffiltext{2}{Department of Astronomy, University of Washington, Box 351580,
Seattle, WA 98195}
\altaffiltext{2}{Astronomy Department, 601 Campbell Hall, University of California, Berkeley, CA., 94720-3411}
\altaffiltext{3}{MIT Kavli Institute for Astrophysics and Space Research, 77
  Massachusetts Ave, 37-582c, Cambridge, MA 02139-4307}
\altaffiltext{4}{Department of Astronomy, Boston University, 725 Commonwealth Ave, Boston, MA 02215}
\altaffiltext{5}{European Southern Observatory, Alonso de Cordova 3107, Casilla 19001, Vitacura, Santiago 19, Chile}
\altaffiltext{6}{Cardiff School of Physics and Astronomy, Cardiff University, Queens Buildings, The Parade, Cardiff, CF24 3AA, UK}
\altaffiltext{7}{UCO/Lick Observatory, Department of Astronomy and Astrophysics, University of California, Santa Cruz, CA 95064}
\altaffiltext{8}{Dept. of Physics and Astronomy, 4129 Frederick Reines Hall,
University of California, Irvine, CA 92697-4575}
\altaffiltext{9}{Apache Point Observatory, P.O. Box 59, Sunspot, NM 88349}

\begin{abstract}
  We present the optical data for 195 HI-selected galaxies that fall
  within both the Sloan Digital Sky Survey (SDSS) and the Parkes
  Equatorial Survey (ES).  The photometric quantities have been
  independently recomputed for our sample using a new photometric
  pipeline optimized for large galaxies, thus correcting for SDSS's
  limited reliability for automatic photometry of angularly large or
  low surface brightness (LSB) galaxies.  We outline the magnitude of
  the uncertainty in the SDSS catalog-level photometry and derive a
  quantitative method for correcting the over-sky subtraction in the
  SDSS photometric pipeline. The main thrust of this paper is to
  present the ES/SDSS sample and discuss the methods behind the
  improved photometry, which will be used in future scientific
  analysis.  We present the overall optical properties of the sample
  and briefly compare to a volume-limited, optically-selected sample.
  Compared to the optically-selected SDSS sample (in the similar
  volume), HI-selected galaxies are bluer and more luminous (fewer
  dwarf ellipticals and more star formation).  However, compared to
  typical SDSS galaxy studies, which have their own selection effect,
  our sample is bluer, fainter and less massive.
\end{abstract}


\section{Introduction}

Galaxies in the local universe span a range of star formation
histories.  At the beginning of the spectrum are gas-rich, low surface
brightness (LSB) galaxies that either have just begun the process of
star formation or have been processing their gas with extremely low
efficiencies.  At the other end are gas-poor galaxies that typically
have formed the bulk of their stars in the distant past and are
currently devoid of gas (Roberts 1963; McGaugh \& de Blok 1997).

Creating a sample that bridges these two regimes requires the union of
two different methods of identifying galaxies.  Stars dominate the
visible light output of most galaxies, and thus galaxies detected by
traditional optical imaging have well developed stellar populations.
In contrast, the natural way to identify gas-rich, less evolved
galaxies is by their 21 cm radio emission. Selecting on HI reveals
galaxies entirely based on their gas content, independent of their
starlight, and thus easily finds systems with intact gas reservoirs
(e.g. Rosenberg \& Schneider 2000).  Characterizing the stellar and
gaseous properties of galaxies selected in the radio and in the
optical will therefore yield information about the entire continuum of
galaxy star formation histories (Burkholder, Impey \& Sprayberry
2001).

Aside from its importance for global star formation, a sample of
gaseous and stellar information for galaxies in the nearby universe
also allows for a more complete census of the local baryons
(Rosenberg, Schneider \& Posson-Brown 2005).  Too often, galaxy
studies neglect the fact that HI dominates the baryonic content of
many galaxies, particularly those with low masses.  The baryonic
makeup of the nearby universe puts important observational constraints
on simulations of galaxy formation and evolution as well as revealing
reservoirs of mass that were previously undetected because of their
optical LSB nature (Disney 1976).  The HI data also provide kinematic
constraints on the dark matter content of the galaxies (Blanton, Geha
\& West 2008).  Therefore, with both optical and HI information, we
can probe how the baryonic content relates to total mass of the
galaxy.

We present the first step toward an inventory of the HI and optical
properties of nearby galaxies.  Our study focuses on an HI-selected
sample and therefore identifies many systems that have retained much
of their primordial HI.  It lacks the systems that have used their
entire gas supply and are dominated by stars.  A separate project is
underway to complete the nearby baryonic census by filling in the gas
poor systems with HI observations of optically selected galaxies.

This study combines data from two high quality, uniform surveys, the
Parkes Equatorial Survey (ES; Garcia-Appadoo et al. 2009), and the
Sloan Digital Sky Survey (SDSS; York et al. 2000; Gunn et al. 1998;
Fukugita et al. 1996; Hogg et al. 2001; Smith et al. 2002; Stoughton
et al. 2002; Pier et al. 2003; Ivezi{\'c} et al. 2004; Gunn et
al. 2006; Tucker et al. 2006). The combination of these two surveys
creates a rich compendium of stellar and HI parameters for galaxies at
various evolutionary states that will be used in future papers to
explore how global star formation proceeds in galaxies as a function
of their physical parameters.

Previously, most HI surveys were targeted at specific locations,
thus little was known about the distribution of HI in the Universe,
independent of optical properties. Henning (1992; 1995) used the NRAO
300 ft telescope to conduct an HI blind survey and recovered 39
sources.  Large blind surveys followed using the Arecibo 300m
telescope, yielding hundreds of sources and allowing for the first
statistically sound studies of the HI mass function (Zwaan et
al. 1997; Spitzak \& Schneider 1998; Rosenberg \& Schneider 2000;
Rosenberg \& Schneider 2002).  These blind surveys also identified
many un-cataloged LSB galaxies and paved the way for more complete
studies of the baryonic content of nearby galaxies (Rosenberg,
Schneider \& Posson-Brown 2005).

Recent studies have combined large HI surveys with optical and
infrared samples, namely the Arecibo Duel Beam and Slice Surveys with
the Two Micron All Sky Survey (2MASS; Jarrett et al. 2000; Rosenberg
et al. 2005), the merging of the HI Parkes All Sky Survey (HIPASS) with SuperCOSMOS (Hambly et
al. 2001a; Hambly, Irwin \& MacGillivary 2001b; Hambly et al. 2001c;
Doyle et al. 2005), and the combination of HIPASS with DSS/POSSII data (Wong et al. 2009). Rosenberg et al. (2005) were able to probe the
baryonic content of a large sample of galaxies, but were limited by
the shallow depth of 2MASS, which does not have data for many of the
LSB galaxies in the sample. The HIPASS/SuperCOSMOS/DSS samples of Doyle et
al. (2005) and Wong et al. (2009) contains optical/IR data for several thousand HI selected galaxies but also suffer from the shallow depth of the SuperCOSMOS/DSS optical data.

\subsection{The Need for a Uniform Sample}
Many studies have investigated the relationships between gas and stars
in galaxies (e.g., Roberts 1963; Fisher \& Tully 1981; Scodeggio \&
Gavazzi 1993; Kennicutt, Tablyn \& Condon 1994; McGaugh \& de Blok
1997; Haynes et al. 1999; Burkholder, Impey \& Sprayberry 2001;
Swaters et al. 2002; Iglesias-Paramo et al. 2003; Karachentsev et
al. 2004; Helmboldt et al. 2004; Rosenberg, et al. 2005; Serra et
al. 2007; Walter et al. 2008; Disney et al. 2008; Zhang et al. 2009).
However, many of these have relied on small inhomogeneous samples.
These studies have been sufficient to establish the broad trend of
increasing gas-richness in low mass systems, but they are limited in
their ability to constrain more accurate relationships between gas,
stars and galaxy mass, as well as the intrinsic scatter in these
physical quantities.  The advent of large astronomical surveys allows
for unions of these large surveys to yield multi-wavelength, uniform
data sets with small systemic errors and large sample sizes
(e.g. Salim et al. 2005; Ag{\" u}eros et al. 2005; Obri{\'c} et
al. 2006; Covey et al. 2008).

As mentioned above, several studies have combined large HI surveys
with large optical and infrared datasets.  In the first of these,
Rosenberg et al. (2005) investigated how the infrared stellar light
compares to the HI gas emission.  Although Rosenberg et al. (2005)
were able to probe the baryonic content of a large sample of galaxies,
they were limited by the shallow depth of 2MASS, which does not have
data for many of the LSB galaxies in the sample.  Their study
therefore excludes the galaxies at the extreme gas-rich end of the
evolutionary spectrum.  Another large-scale blind HI survey is the HI
Parkes All Sky Survey (HIPASS; Barnes et al. 2001; Meyer et al. 2004;
Zwaan et al. 2004; Wong et al. 2006), which has been combined with IR
and optical catalogs (Doyle et al. 2005; Wong et al. 2009) that are
very similar to our ES/SDSS catalog (see below).  In fact, a large
fraction of the ES data is in the HIPASS catalog (see Garcia-Appadoo
et al. 2009 for more information).  While the HIPASS studies have much
more sky coverage than our ES/SDSS catalog, the depth of the SDSS
allows us to probe optical magnitudes several times fainter and
recover optical counterparts for all of the HI sources in our
footprint.


The union of SDSS and the ES provides the desired uniformity in both
the optical and the radio (HI) data, along with remarkable depth and
dynamic range of the optical SDSS data.  Although there is only a
modest area of SDSS/ES overlap, enough data exist for the
construction of a uniform HI-selected catalog that can be used to
probe how the baryonic content of galaxies changes as a function of
other physical parameters.

This paper is one of several papers utilizing the combined
ES/SDSS data.  In this paper we describe the sample selection, discuss
the methods by which we derive the optical photometric parameters, and
present the optical data. We briefly describe the sample
characteristics and compare our sample to an optically selected sample
in a similar volume.  Other papers describe the HI data
(Garcia-Appadoo et al. 2009) and explore the gas fractions, colors (West et al. 2009), and dynamics of the ES/SDSS galaxies.

\section{Survey Descriptions and Sample Selection}
\subsection{Equatorial Survey}

The ES was carried out with the Parkes Multibeam system on the 64m
radio telescope in Parkes, Australia (Stavely-Smith et al. 1996).  The
ES, which is described in detail in an accompanying paper
(Garcia-Appadoo et al. 2009), circles the celestial equator between -6
$< \delta <$ +10 and contains over 1000 sources in 5738 square
degrees.  The raw data forms part of HIPASS (Barnes et al. 2001; Meyer
et al. 2004; Zwaan et al. 2004; Wong et al. 2006), carried out with
the same instrument over the entire sky between -90 $< \delta <$ +25.
However, the ES fields were searched much earlier (Garcia-Appadoo et
al. 2009) in readiness for comparison with the earliest SDSS data.
While the search techniques were much the same as the HIPASS team's
and rely heavily on their procedures, the source lists are not
identical.  For example, the completeness limit of the ES list is 30\%
fainter than the HIPASS limit.  This is mainly due to our ability to
follow-up and confirm a higher proportion of the fainter sources, a
process that would be impractical with the larger
survey. Garcia-Appadoo et al. (2009) includes a more detailed
comparison between HIPASS and the ES.

The ES covers a velocity range from -1280 to 12700 km\ s$^{-1}$ with an
RMS noise of 13 mJy.  The velocity resolution of the ES HI spectra is
18.0 km\ s$^{-1}$ and the 3$\sigma$ HI mass limit of the survey is
$10^6\times D^2_{Mpc}M_{\odot}$, assuming a 200 km\ s$^{-1}$ HI galaxy
profile.  For detailed descriptions of the data acquisition,
calibration and reliability see Garcia-Appadoo et al. (2009) and
the HIPASS analysis contained in Barnes et al. (2001), Meyer et
al. (2004), and Zwaan et al. (2004).

The ES data cubes were searched using an automated search code written
in {\tt MIRIAD}.  1164 sources were extracted from regions of the sky
where SDSS overlaps were likely to occur.  For each source detection,
the HI spectrum was extracted by fitting a baseline to the background
flux (Barnes et al. 2001).  The source position, recessional velocity,
20\% peak velocity width, peak and integrated fluxes were measured
from the spectrum.  For further details on the HI source extraction
and parameter measurements for the ES-SDSS sample, see the
description in Garcia-Appadoo (2009).

\subsection{SDSS Survey}
The optical data for this study come from the SDSS Data Release 2
(DR2; Abazajian et al. 2004) sky area.  The DR2 area is 3324 deg$^2$,
$\sim$1700 deg$^2$ of which overlaps with the 5738 square degree ES region discussed
above.  The majority of the overlap falls along the $\delta$=0 strip
(excluding the Galactic plane) and the DR2 coverage of the northern
Galactic cap (-5$\lesssim \delta \lesssim$5; see Garcia-Appadoo et al. 2009 for further details).  Automatic pipelines
(Pier et al. 2003; Lupton et al. 2002) reduce the raw data and store
the derived quantities in a catalog.  The SDSS photometric pipeline
({\tt PHOTO}; Lupton et al. 2002) is optimized for speed and the faint
end of galaxy population. Galaxies in the tail of angular size
distribution (large) and surface brightness distribution (low) often
have unreliable measurements (see Section 2.3).  Therefore, aside
from the initial catalog matching described below, no SDSS catalog
data were used for this study.  All photometric quantities were
obtained using a new set of techniques optimized for deriving large
galaxy photometry.

\subsection{Deblending and Sky Subtraction in SDSS}

The SDSS automatic pipelines were optimized for angularly small, faint objects.  It was found that the SDSS catalog values for angularly large, bright galaxies are often unreliable.  Before we re-derive the photometric values for the ES/SDSS sample, we will briefly discuss and quantify two of the problems with large, bright galaxies in the SDSS, namely issues with deblending (or shredding), and sky subtraction.  Both of these problems have been discussed in previous papers (e.g. Abazajian et al. 2004, 2009) but require additional attention as they have particular relevance for the ES/SDSS sample.  

The SDSS photometric pipeline ({\tt PHOTO}; Lupton et al. 2002) identifies objects in SDSS fields and extracts them into individual atlas images for photometric analysis.  When one objects falls in front (or near) another on an SDSS field, the {\tt parent} images is sent to the deblender.  The deblender separates the two objects into {\tt children} and photometry is performed on both images independently.  For most objects in SDSS, this process works well.  However, in galaxies with large angular extent, irregular morphology or low-surface brightness, {\tt PHOTO} deblends more {\tt children} than it should; HII regions and spiral arms are frequently separated from their {\tt parent} galaxies. Because these galaxies are ``shredded'' into multiple pieces, accurate photometry is prohibited by the standard photometric pipeline.  

\begin{figure}
\centering
\plotone{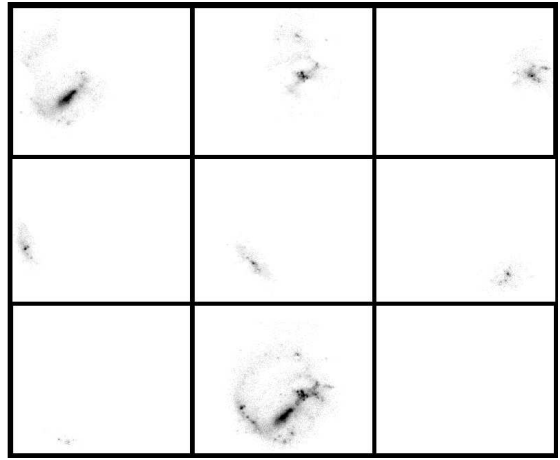}
\caption{$r$-band atlas images for HIPEQ1124+03.  The deblender has divided this galaxy into 7 children.  The ``brightest child'' (upper left) contains only 50\% of the total galaxy flux.  The irregular morphology of this system is responsible for the large degree of ``shredding'.''  The combined image is in the lower-middle panel.}
\label{deblend}
\end{figure}

\begin{figure}
\centering
\plotone{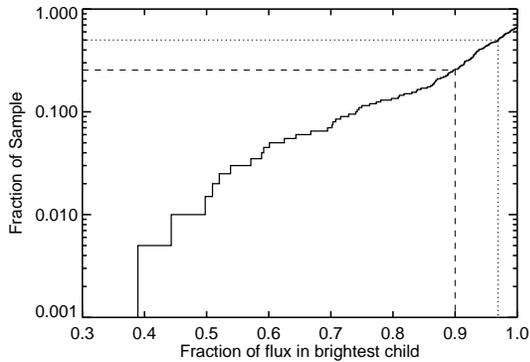}
\caption{Cumulative distribution of the fraction of flux in the brightest child of deblended ES/SDSS galaxies.  The dashed line indicates the level at which galaxies have at least 90\% of their flux contained in the brightest child.  This accounts for more than 75\% of the total sample.  The dotted line shows the median value of the sample.  Half of the objects have brightest children with more than 96\% of the total flux.}
\label{distphot}
\end{figure}

Using the ES/SDSS sample, we demonstrate the amplitude of the deblending problems in the SDSS catalog.  Some of the galaxies deblend perfectly; no significant amount of flux has been removed and the foreground stars have been correctly deblended.  However, in some cases, the deblending is quite extreme.  Figure \ref{deblend} shows the 7 {\tt children} that were shredded off of the ES/SDSS source HIPEQ1124+03.  The aggressive deblending resulted in the brightest {\tt child} (upper left) only containing 50\% of the of total flux.  The SDSS catalog photometry for HIPEQ1124+03 is therefore completely unreliable.  We quantified the magnitude of the deblending problem in the ES/SDSS sample by examining the fraction of flux in the brightest {\tt child} for each of the ES/SDSS galaxies.  Figure \ref{distphot} shows a cumulative distribution of of the fraction of flux contained in the brightest {\tt child} for the ES/SDSS sample.  Roughly 75\% of the the galaxies have more than 90\% of their flux contained in the brightest {\tt child}.  The remaining 25\% have irregular morphologies, are flocculant (HII regions are removed as stars) or have low surface brightness, with a number of brightest {\tt children} having less than 50\% of the total galaxy flux.  We discuss our method for remedying the deblending problems in the ES/SDSS sample in Section 4.1.

\begin{figure}
\centering
\epsscale{0.6}
\plotone{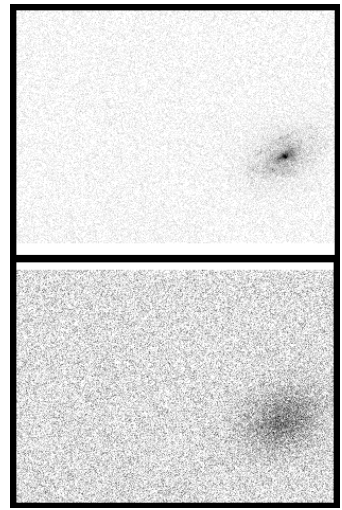}
\caption{Reconstructed Field for HIPEQ1232+00a with only the relevant atlas images included in the field (top).  Rebuilt {\tt{PHOTO}} sky for galaxy HIPEQ1232+00a (bottom).  The bottom field is what {\tt{PHOTO}} subtracted from the corrected frame before photometry was performed.  Flux from the galaxy is clearly seen on this image, highlighting the overestimation of sky for large, bright galaxies.}
\label{skyex}
\end{figure}

Problems with the sky subtraction for bright SDSS galaxies have been recently identified in the literature (Mandelbaum et al. 2005; Bernardi 2007; Lauer et al. 2007; Adelman-McCarthy et al. 2008; Abazajian et al. 2009).  Previous simulations indicate that the magnitude of the sky subtraction error can be as high as 0.4 magnitudes (Masjedi et al. 2006; Abazajian et al. 2009). Like deblending, the sky subtraction algorithm (see Lupton et al. 2002) is optimized for small, faint objects and using a 256 $\times$ 256 pixel mask to determine the sky value every 128 pixels in an SDSS field.  If a galaxy in the field takes up an appreciable fraction of the mask, the sky is overestimated (and over-subtracted).  Therefore, we find that angular size, and not magnitude is the main determiner of sky subtraction error.  Figure \ref{skyex} shows a reconstructed SDSS field (top; see Section 4.1) and the rebuilt sky (bottom) for ES/SDSS source HIPEQ1232+00a.  An accurate sky determination should yield a smooth surface.  However, the bottom panel clearly contains flux from the galaxy, which results in fainter magnitudes when it is subtracted from the top image.

\begin{figure}
\centering
\epsscale{1}
\plotone{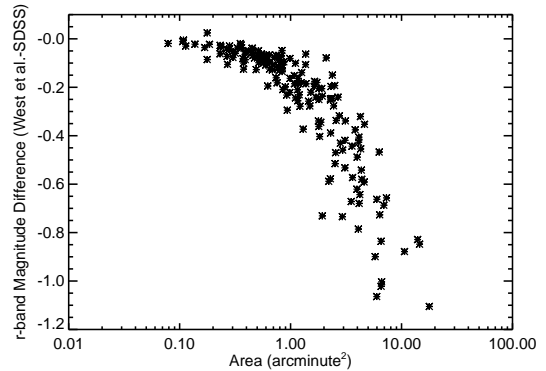}
\caption{Difference between the $r$-band magnitude derived using our sky subtraction pipeline and that of SDSS as a function of the area with the PetroR90 radius for the ES/SDSS sample.  For galaxies with areas larger than 0.5 $\square^{\prime}$, the loss from bad sky subtraction is substantial.}
\label{skymag}
\end{figure}

To investigate the magnitude of the sky subtraction problem for the ES/SDSS sample, we ran our photometric software (see Section 4.4) on the ES/SDSS galaxies with both the standard SDSS sky subtraction and our sky determination (see Section 4.2) and compared the results.  Figure \ref{skymag} shows the difference in $r$-band magnitude between the 2 sky subtraction methods for all of the ES/SDSS galaxies as a function of the area of the galaxy (determined at the 90\% light radius).  While the effect is negligible for small galaxies ($<$ 0.5 $\square^{\prime}$), galaxies with angular areas of $\sim$1 $\square^{\prime}$ have magnitude errors greater than 0.2 magnitudes.  There are several galaxies in the ES/SDSS sample where the magnitude error from sky subtraction is larger than 1 magnitude.

The scatter in Fig. \ref{skymag} implies that other parameters beyond angular size affect the flux lost to sky subtraction errors.  Instead of looking at the magnitude difference (which is fractional flux), we explored the actual flux lost as a function of galaxy size, magnitude and axis ratio.  The larger galaxies fill more of the sky mask, resulting in more of their flux being subtracted.  But the amount of flux depends on the surface brightness of the source (which depends on magnitude, area and axis ratio).  

\begin{figure}
\centering
\plotone{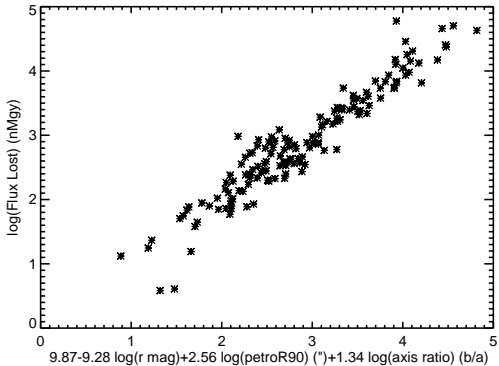}
\caption{Projection of the best fit hyper-plane to log($m_r$), log(petroR90), log(axis ratio) and log(flux lost) for the ES/SDSS galaxies.  The scatter is less than half of an order of magnitude.  The same relation applies to each of the other 4 bandpasses.}  
\label{skyfit}
\end{figure}

We fit a hyper-plane to the logarithms of the flux lost (in nanomaggies\footnote{Nanomaggies (nMgy) are a standardized flux unit where   $m$=22.5-2.5$\log$(nMgy)}; Blanton et al. 2003b), the r-band magnitudes , the Petrosian 90\% light radii and the axis ratios (all measured using the SDSS catalog values).  Figure \ref{skyfit} shows the projection of the best-fit hyper-plane for all of the ES/SDSS galaxies and can be described by:

\begin{eqnarray}
\log(flux_{lost}) &= 9.87 -9.28\log(m_r)  & \nonumber\\
  & +2.56\log(petroR90)   & \\
  & +1.34\log(b/a) \pm 0.23 & \nonumber 
\end{eqnarray}

\noindent where the flux is measured in units of nanomaggies, $petroR90$ is the SDSS Petrosian 90\% light radius measured in arcseconds, and $b/a$ is the axis ratio of the galaxy.  The axis ratio ($b/a$) is determined from the major and minor axes derived from SDSS isophotal photometry ({\tt isoA} and {\tt isoB} in {\tt PHOTO} respectively). Future analysis of bright, angularly large galaxies in SDSS can be corrected at the catalog level using Eq. 1.  However, for the ES/SDSS sample, we will use our sky subtraction method that is described below (Section 4.2).

\subsection{Comparison Sample}
To compare the HI-selected ES/SDSS galaxies to an optically
selected sample, we use the Data Release 4 (DR4; Adelman-McCarthy et
al. 2006) ``main'' galaxy sample with $m_{r,Petrosian}$ $<$ 17.77 mag.
(Strauss et al. 2002).  We apply a secondary cut of  $z$ $<$
0.04 to ensure that a small local volume is being sampled.  There are 30,236 SDSS galaxies in this volume that we will use for comparison.  Because
there were no major photometry changes between DR4 and DR2, there is
no concern in comparing the DR4 data of our main comparison sample to
our ES/SDSS sample drawn from DR2.  Because of the surface
brightness limit for SDSS spectroscopic observations (Strauss et
al. 2002), the comparison sample does not contain any galaxies with
$\mu_{r} > 23.0$, where $\mu_{r}$ is the Petrosian surface brightness
(within the 50\% radius).

The comparison sample serves two distinct purposes: 1) to highlight the fact that an HI-selected sample selects a different subset of galaxies than are available in the SDSS spectroscopic catalog; and 2) to compare the properties of an optically-selected sample to those of an HI-selected sample. The latter requires that we constrain our comparisons to galaxies that fall within the same volume. There are 51 ES/SDSS galaxies and 18,379 SDSS ``main'' galaxies in the overlapping volume.

\subsection{Matching/ Confirmation}
\subsubsection{SDSS Photometry}
The catalog matching began by searching the DR2 {\tt tsObj} files for
all SDSS sources within 10$^{\prime}$ of the ES source positions. A
web-page showing the field for each matching SDSS object was created
and all candidate objects were sorted by inverse object size (large to
small) to ease visual inspection.  Over 1.16 million SDSS objects were
found inside the beam areas of 310 ES sources that fell within the SDSS
DR2 footprint.  At every ES source position, the candidate SDSS
objects were visually inspected and potential counterpart galaxies
identified.

To be included in the ES/SDSS sample, each candidate galaxy had to
meet 4 criteria: 1) the ES recessional velocity must agree to
within twice the $W_{20}$ value of the optically derived redshift; 2)
there must be no more than 1 detectable, spatially resolved galaxy
within the ES beam at the same redshift; 3) the candidate galaxy
must not extend across two or more SDSS fields; and 4) all galaxies
must be at least 1$^{\prime}$ away from any saturated foreground
stars.

To apply the
first criterion, we obtained a redshift for each candidate galaxy.
SDSS spectra and redshifts were available for $\sim$80\% of the
candidate galaxies. For the remaining candidates, we searched the
NED\footnote{This research has made use of the NASA/IPAC Extragalactic
  Database (NED) which is operated by the Jet Propulsion Laboratory,
  California Institute of Technology, under contract with the National
  Aeronautics and Space Administration} database and acquired
redshifts for all but $\sim20$ galaxies.  The remaining sources were
spectroscopically observed using the Apache Point Observatory's (APO)
ARC 3.5m telescope from February 2002 to July 2003.  All of the
sources were observed with long integrations ($>20$ minutes) on the
Dual Imaging Spectrograph (DIS) with a 1.5$^{\prime\prime}$ slit and
with the high resolution gratings ($\sim$2\AA).  Most of the galaxies
in the ES/SDSS sample are currently forming stars and have emission
lines that can be unambiguously identified and easily measured for
accurate redshift determination.  Most of
the galaxies have HI and optical recessional velocities that match to
within half of the HI line width (see Fig. 10 of Garcia-Appadoo et al. 2009).  After obtaining redshifts for all of the candidate sources, \emph{all} 310 ES sources in the DR2 footprint had SDSS galaxies within the ES beam at the same redshift. 

Of these 310 galaxies, 90 failed the second criterion due to multiple
SDSS galaxies within a single ES beam.  Some of the HI detections
had as many as 5 galaxies at the same redshift, making it impossible
to assign an HI mass to an individual galaxy.  With multiple galaxies
in the ES beam, only the total HI of the group is measured, and
without higher resolution 21cm observations, this problem cannot be
resolved.  This may introduce a bias against galaxies in higher
density regions when determining the HI mass function from the
combined ES/SDSS sample.  

At the time of sample selection, no techniques were available to
accurately obtain the photometry for galaxies with flux spread over
multiple fields.  Twenty of the galaxies were positioned in such a way
and/or had angular extents so large that they fell over multiple SDSS
fields (criterion \#3) and were therefore excluded from our catalog.
This selection criteria also introduces a bias against the largest
nearby galaxies.

Five additional galaxies were removed because of their close
proximity to saturated foreground stars (criterion \#4) whose scattered light would greatly
effect galaxy photometry.  Since the superposition of stars in front
of galaxies is random, this criterion should not introduce any additional bias.

The resulting ES/SDSS sample consists of the 195 galaxies that passed
all four criteria.  Their survey names, central SDSS positions, other
catalog names, and morphological types from NED can be found in Table
\ref{names}\footnote{The full tables shown in this paper are available
  in the electronic version of the journal.  In addition, all measured
  and derived photometric quantities for the ES/SDSS sample can be
  obtained electronically using the CDS Vizier database
  http://vizier.u-strasbg.fr/viz-bin/VizieR}.  The position centers
are those used for photometry.  We refer to objects both in the text
and in tables by their ES (HIPEQ) catalog name.


\section{Distance}

\begin{figure}
\centering
\plotone{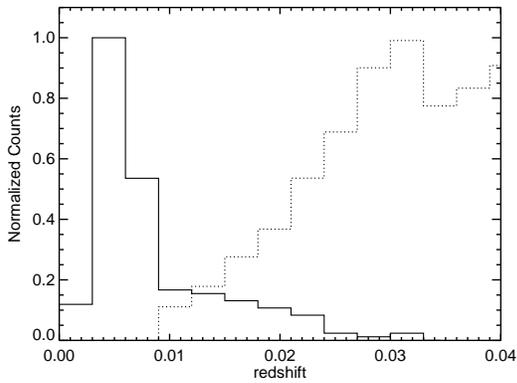}
\caption{The redshift distribution of the HI-selected galaxies (solid) and of a volume-limited sample of SDSS galaxies (dotted) from the Data Release 4 (DR4) main galaxy sample. The HI-selected galaxies are much closer than those in most SDSS galaxy studies. (A similar figure using recessional velocity in place of redshift was included in the Garcia-Appadoo et al. 2009 paper).}
\label{zdist}
\end{figure}

Because of the limits of the ES survey, the HI-selected sample
probes only the very nearby Universe.  Figure \ref{zdist} shows the
redshift distribution of the HI-selected galaxies (solid) with the
redshift distribution of a volume-limited sample of SDSS galaxies
(dotted) from the DR4 main galaxy sample. The galaxies in the
HI-selected sample are clearly biased toward very small distances,
compared to the galaxies included in most SDSS studies.  Although this
means that the HI-selected sample probes a smaller volume than other
SDSS studies, the volume it samples is more complete because it
includes substantial low surface brightness and low luminosity
populations (although the sample is still missing gas-free galaxies
eliminated by the selection criteria).

Many measured recessional velocities are likely to be influenced by
infall toward Virgo.  We adopt a cosmology with H$_0$=70 km\ s$^{-1}$ Mpc$^{-1}$,
$\Omega_M$=0.3 and $\Omega_{\Lambda}$=0.7 and then correct
recessional velocities for Virgo infall using the {\tt IDL} routine
{\tt v\_converter}, which derives velocity corrections from the LEDA
database (Theureau et al. 1998; Terry et al. 2002).  The velocities
are then converted into redshifts and luminosity distance.  The
adopted distances can be found in Table \ref{table:derive}.

\section{HI-Selected Galaxies - Photometry}

SDSS photometry of large galaxies is non-trivial.  In almost every
SDSS galaxy study to date, large, nearby galaxies have been
purposefully avoided because of the challenge in correctly extracting
their photometric quantities.  Problems with deblending and sky
subtraction preclude an automated catalog-level analysis of large
galaxies, and the inclination problems in the Petrosian quantities
derived by SDSS adversely affect the photometry of both large and
small galaxies (Bailin \& Harris 2008).  In this section we discuss the
limitations of SDSS photometry and outline the methods used to remedy
the situation.  We then apply these techniques to derive photometry
for the ES/SDSS sample.

\subsection{Creating New Atlas Images}

\subsubsection{Reconstruction of Atlas Images}


For each galaxy in the ES/SDSS sample, we downloaded all atlas images within
a 5$^{\prime}$ radius of the galaxy center.  We visually inspected
each atlas image and identified the children that belonged to the
ES/SDSS galaxy.  For the most part, the appropriate children were
easy to identify because of their uniform colors and extended
morphology.  In a few cases however, confusion between an HII region
and a foreground star made the choice difficult.  In these cases, we
examined the colors of the star-like objects and included them as
children if they fell off the stellar locus as defined by Covey et
al. (2007).  The {\tt{object\_id}} for likely parts of the galaxies
were saved, and the corresponding images were mosaicked together to
reconstruct an image uncontaminated by stars.  The reconstructed image
was visually inspected to ensure that it did not have significant
missing children and that bright stars and their artifacts were not
included in the reconstructed images.

To properly reconstruct the images of the galaxies with their original
sky, we used the {\tt tsObj}, {\tt fpAtlas}, {\tt fpBIN} and {\tt
fpFieldStat} SDSS files to rebuild corrected frames ({\tt fpC} files)
from a specified list of atlas images.  The resulting images contain
only the galaxy superimposed on a restored image of the sky.  We also reconstructed single galaxy {\tt fpC} files and sky images for every field.  The latter were useful in
correcting for the error in the SDSS sky subtraction (see Section 4.2
below).

Finally, we aligned all five SDSS bands to the same pixel coordinates
and made cut-outs of each galaxy while preserving the astrometric
(and other header) information.

\subsubsection{Masking Out Stars}
Twelve of the galaxies in the sample were so badly deblended by {\tt
PHOTO} that they could not be easily reconstructed from their
deblended children.  Instead, we created a sky-subtracted {\tt fpC}
file of the deblended parent galaxy.  To remove the stars from the
atlas images, we masked out the bright stars and filled in the masked
regions by interpolating from surrounding pixels.

\subsection{Sky Subtraction}
\subsubsection{Sky Subtraction of SDSS Fields}
As outlined above, sky subtraction errors for large
galaxies can be larger than 1 magnitude in the SDSS catalog data.
These errors are significant and must be avoided for the large
galaxies (area $>0.5 \Box^{\prime}$) that dominate our sample.

To subtract the sky from our images, we followed the procedure
outlined below.  We ran {\tt SExtractor} (Bertin \& Arnouts 1996) on
an $r$-band subregion of a corrected frame({\tt fpC} file) to identify
sources and masked all non-sky pixels, growing masks where needed to
eliminate all galaxy flux from sky regions.  Then we fit a tilted
plane to the remaining sky pixels and subtracted the tilted plane fit
from the reconstructed galaxy image.  We repeated this procedure for
all 5 bands ($u$, $g$, $r$, $i$, $z$) using the masks derived from the
$r$-band subregion.

\subsubsection{Verifying Sky Subtraction Accuracy}
Many of the sources in our catalog have very low surface brightness,
making their photometry particularly sensitive to errors in sky
subtraction.  Unlike random errors due to dark current and Poisson
noise, errors in sky determination lead to systematic over or
under-subtraction of flux from every pixel in a galaxy image.  Thus,
errors in the sky level do not average to zero over a galaxy image,
and instead can be a major source of uncertainty, particularly for
large and/or low surface brightness galaxies.  In this section we
explore various methods of calculating the sky background and show
that almost all of the methods return similar results.  We used the
differences among sky subtraction techniques to characterize the
uncertainty in determining the sky background.

We selected a sparsely populated field from the SDSS on which to test
various sky subtraction methods.  We tested twelve different methods
using the combination of three different sky field cutouts and four
different parameterizations for the shape of the sky background
surface fits, ranging from first-order (i.e. a tilted plane) to
fourth-order (i.e. a fourth order polynomial in x and y) respectively.
The different sky fields are: 1) an SDSS field
(13$^{\prime}\times9^{\prime}$) with all of the objects masked out; 2)
an SDSS field with all of the objects masked out and the mask of
HIPEQ0014-00 (a medium/large sized galaxy in the sample with an area
of 1.4 arcmin$^2$) superimposed on the field.  This additional masking
leaves a large ``hole'' in the image at the galaxy location, adding
uncertainty to the fitting procedure; and 3) a
4$^{\prime}\times5^{\prime}$ subregion of an SDSS field with all of
the objects masked out and the mask of HIPEQ0014-00 superimposed on
the cutout.  Each of the twelve methods was run on all 5 bands of the
sparsely populated SDSS test image.

To quantify the differences among the various methods, we calculated
the average residual counts per pixel in the area under the
HIPEQ0014-00 mask corresponding to the area where galaxy photometry
would be measured. We assumed that a perfect sky subtraction yields a
mean residual sky level of zero.  Thus, any non-zero mean sky level is a
measure of the sky subtraction uncertainty, and is reported as the
residual.  This analysis adequately reproduces the manner in which
actual sky subtraction of a galaxy field is calculated and assesses
the relevant residuals that affect galaxy photometry.  

The maximum residual in any of the 5 bands is 0.1 counts/pixel and the
RMS is 0.04 counts/pixel. For an $m_r$ $\sim13$ galaxy with an aperture
area of 4$\Box^{\prime}$, a residual of 0.1 count/pixel results in a
0.01$^{\rm{m}}$ change.  For comparison, a residual of 1 count/pixel
would result in a 0.2$^{\rm{m}}$ change for the same galaxy.  The
residuals show no clear trend with either the fit order or the area
schema.  However, the $i$-band data have systematically larger
residuals while the $z$-band's are systematically smaller.  This may
be due to faint red objects that are unidentified in the $r$ band,
detected in $i$ and marginally detected in $z$ (which is substantially
shallower than $i$).  These objects would not be properly removed
using the $r$-band masks.

All of the methods have residuals are that well below other per pixel
uncertainties that will be described below (e.g. the {\tt Dark
Variance} varies between 0.9 and 3.9 counts/pixel). Therefore, we
conclude that choosing any of these methods will not have an effect of
more than $\sim0.01^{\rm{m}}$ on the resulting photometry.

For the photometry presented in this study, we have adopted a tilted
plane fit to a subregion of the SDSS field for the sky
subtraction. This approach is the computationally fastest method and,
although the $z$-band residual is slightly worse than other methods,
the deviation is still not significant.  For completeness we use the
largest absolute residual in each band (regardless of method) as a
measure of the uncertainty in the sky value.  This is added to the
uncertainty analysis described below.

We note that the SDSS residual (defined as the mean difference between
the SDSS sky and the method presented in this paper) for ES/SDSS
galaxy HIPEQ0014-00 is $\sim$1.13 counts/pixel, demonstrating the
improvement made over the SDSS pipeline sky subtraction.  The improved
sky subtraction changes the $r$-band magnitude of HIPEQ0014-00 by 0.19
magnitudes.

\subsection{Model Fitting}
The reconstructed sky subtracted galaxy images were run through two
different model fitting routines to quantify the galaxies' sizes,
surface brightnesses, and orientations.  Both fitting routines use the
Levenberg-Marquardt (LM) minimization technique to calculate a
two-dimensional surface fit to a galaxy image.  The first routine fits
a single S\'{e}rsic (1968) profile and the second fits a two
component, double exponential disk.  All fits were performed on the
$r$-band images.  If the fit was able to accurately reproduce the
galaxy image, then an elliptical aperture was used for subsequent
photometry.  47 of the galaxies were not well fit by this simple
profile because of their irregular morphologies and/or low surface
brightnesses.  For the irregular and LSB galaxies the parameters are
not reliable and a default circular aperture is used for photometry.
Although the lack of seeing in the models may introduce a small amount
of uncertainty into the exact S\'{e}rsic profile derived, it has
almost no effect on the aperture shape

\subsection{Petrosian Photometry}

In this section, we discuss the derivation of  Petrosian
quantities for the ES/SDSS sample.  Petrosian (1976) photometry
recovers a nearly constant fraction of a galaxy's flux for a variety
of morphological types and surface brightness profiles.  The resulting
photometry has fewer biases than those that estimate the total galaxy
flux with apertures based on isophotes or fractions of the central
surface brightness.  A modified Petrosian photometric system has been
adopted by SDSS (see Blanton et al. 2001; Yasuda et al. 2001).

Petrosian quantities were adopted for this study to be consistent with
SDSS photometry, allowing us to compare our results to the greater
SDSS sample.  The SDSS photometric pipeline uses circular apertures to
extract Petrosian quantities.  However, this produces a significant
inclination dependence in some of the resulting photometric quantities
(Bailin \& Harris 2008).  This bias is worse in our sample than the Main SDSS sample because the ES/SDSS galaxies are sufficiently nearby that their inclinations are unaffected by seeing (which tends to circularize angularly small galaxies).  We have remedied this problem by allowing for
elliptical apertures in our Petrosian photometric pipeline.  For the
galaxies with high quality S\'{e}rsic fits, we adapted the SDSS
photometric methods to include elliptically shaped photometric
apertures.  We also measured every galaxy with a circular aperture for
comparison with standard SDSS outputs.  These circular measurements
were adopted for galaxies whose S\'{e}rsic fits were of poor quality.


The first step in deriving Petrosian quantities was to calculate the
Petrosian radius.  Following the SDSS prescription, we defined the
Petrosian ratio $R_P$ as:

\begin{equation}
R_{P}(r)\equiv\frac{\int_{0.8r}^{1.25r} dr^{'} 2 \pi r^{'} I(r^{'})/[\pi (1.25^2-0.8^2)r^2]}{\int_{0}^{r} dr^{'} 2 \pi r^{'} I(r^{'})/(\pi r^2)} 
\end{equation}

\noindent where $I(r)$ is the azimuthally averaged surface brightness
profile.  We performed the integration by taking 1 pixel
steps and calculating the subsequent Petrosian ratio at every pixel from
the galaxy center.  In the SDSS pipeline, the Petrosian radius $r_{P}$
is defined as the radius at which $R_{P}(r)$ is equal to 0.2 (see
Blanton et al. 2001, Yasuda et al. 2001).  The Petrosian flux $f_{P}$
in any band was then defined as the flux within 2.0 Petrosian radii:

\begin{equation}
f_{P}\equiv\int_{0}^{2r_{P}}2\pi r^{'} I(r^{'})dr^{'}
\end{equation}

Our method differs from the SDSS method in that we use elliptical
apertures, in addition to
the standard SDSS circular apertures.

To correctly measure colors of the galaxies, the magnitude
determination for all five bands must use the same aperture. Mimicking
SDSS, we used the derived $r$-band aperture for all bands.  

In some cases the Petrosian radius was so large that the derived
aperture ($r_P$) far exceeded the observed boundaries of the galaxy.
In these cases, large amounts of sky pixels were included in the
galaxy flux.  Although, in the ideal case, the summation of all sky
pixels is zero, we know from Section 4.2 that small residuals exist, and
thus, as Petrosian radius increases, so does the error from the sky
subtraction residuals. Due to their uncertain, irregular surface
brightness profiles, most of the cases where the Petrosian radius was
exceedingly large were LSB galaxies.  The resulting flux errors were
particularly large for these systems because the integrated sky
subtraction residuals can be a large fraction of the total galaxy
flux. A similar problem is present in the SDSS photometric pipeline
(Lupton, private communication).  SDSS corrects for this by using the
statistically defined edges of atlas images as the maximum possible
photometric boundaries.  We imposed a similar maximum size for the
Petrosian radius using the masks created in sky subtraction for each
$r$-band image as the photometric boundaries (since the aperture is
applied to all 5 bands, this does not affect the measured color of the
galaxies).  For most of the galaxies, the integration of Petrosian
magnitudes rarely reached the mask boundaries, indicating that the
galaxy flux was correctly masked out in the sky determination.  The
only galaxies that reached the mask boundary were the very LSB
galaxies, whose Petrosian radii were not well defined (29\% of the
ES/SDSS sample).  Typically, the resulting flux errors from this were
only 1-2\% and were propagated into the reported magnitude
uncertainties.

Once a Petrosian flux was calculated, we converted all fluxes to asinh
magnitudes (Lupton, Gunn \& Szalay 1999) using the relation:

\begin{equation}
mag = -\frac{2.5}{\ln(10)}\left[{\rm{asinh}}\left(\left(\frac{f}{f_0}\right)/2b\right)+{\ln(b)}\right]
\end{equation}

\noindent where $b$ is the softening parameter (a constant for each filter) and $\frac{f}{f_{0}}$ is given by:

\begin{equation}
\frac{f}{f_{0}} = \frac{counts}{exptime} 10^{0.4(aa + kk\times airmass)}.
\end{equation}

\noindent Here the $aa$ is the magnitude zeropoint, measured for each
field, $kk$ is the extinction coefficient in magnitudes, and $exptime$
is the standard SDSS value of 53.907456 seconds.  The values for
$airmass$, $aa$, and $kk$ were all pulled from the corresponding {\tt
  tsField} files.  Asinh magnitudes were computed for both the
elliptical and circular apertures. The resulting magnitudes can be
found in Table \ref{table:paper}. If a galaxy's profile was not well
fit by an elliptical aperture, then the Petrosian circular aperture magnitude
was included in Table \ref{table:paper}.

For most of the galaxies in the ES/SDSS sample, the main
contribution to the photometric uncertainty was dominated by the
Poisson noise of the subtracted sky. The Poisson variable for CCD
photometry is $photo\_electrons$.  We converted from counts to
photo\_electrons using the following relation:

\begin{equation}
photo\_electrons = counts\times gain
\end{equation}

\noindent where the $gain$ is a variable quantity that is also stored for each filter in each {\tt field} in the {\tt tsField} files.

The photometric uncertainty was then derived to be:

\begin{equation}
\sigma(cts) = \sqrt{\frac{cts+sky}{gain} + N_{pix}(dark\_var+skyErr)}
\end{equation}

\noindent where $cts$ is the galaxy flux in counts, $sky$ is the
integrated sky flux in the object aperture, $N_{pix}$ is the number of
pixels in the object aperture, $dark\_var$ is a combination of
the dark current and the read noise per pixel that was found in the
{\tt tsField} files and significantly varies from field to field and
filter to filter.  The $skyErr$ term is the maximum residual from sky
subtraction in each of the five band passes.  The uncertainties from
Eq. 7 (in counts) were then converted to magnitudes using Eq.
4.  In most cases the $counts$ term dominates the total uncertainty.

\subsection{Photometric Corrections}
Variations in extinction, cosmic redshift and SDSS field boundaries
can lead to significant offsets in the observed colors and magnitudes
of galaxies. Thus, various photometric corrections need to be made to
responsibly use the aforementioned photometry for science.  In this
section we outline the photometric corrections that are important for the ES/SDSS sample.

\subsubsection{Edge Corrections}

Of the 195 galaxies in the HI-selected sample, 34 galaxies are close
to the edge of an SDSS field, but were included in the sample because
the majority of their light falls within the SDSS field. The flux lost
beyond the edge boundary should not drastically affect the color of a
galaxy, but it can change the derived absolute magnitude.  The
S\'{e}rsic fit models can be used to make a reasonable estimate of the
flux lost across the field edge.

To correct for the missing flux we overlaid the S\'{e}rsic models for
each of the 34 galaxies and masked out any regions that overlapped (leaving
only model flux that crosses the SDSS field boundaries). Using the
same Petrosian apertures described in Section 4.4, we calculated the flux remaining from the S\'{e}rsic models and converted this
additional flux to a magnitude correction. Only 23 of the 34 galaxies
had any significant flux ($\geq$ 0.01 magnitudes) that was lost
across the field edge and most of the corrections are smaller than 0.1 magnitudes.  Because the S\'{e}rsic models are only
calculated in the $r$-band, we assume that a similar fraction of flux
is lost in the other bands and use the $r$-band value as a universal
magnitude correction.  This assumption is reasonable for small
corrections but may be incorrect for larger amounts of lost flux in
galaxies with large color gradients.  The values for the edge
correction are included in Table \ref{table:photcor}.


\subsubsection{Extinction}

The photometric magnitudes listed in Table \ref{table:paper} are not
corrected for either extinction from the Milky Way or internal
extinction from the extragalactic object itself.  The former is
provided by Schlegel, Finkbeiner, \& Davis (1998; SFD), using dust maps of
the Milky Way.  The SDSS database provides the extinction values in
all five bands at every SDSS pointing. The SFD extinction values are given in Table \ref{table:photcor}.

The correction for a galaxy's internal extinction is more complicated.  The
literature gives examples of extinction corrections that have been
applied on the basis of Hubble type (Gavazzi \& Boselli 1996), rotation
speed (Tully et al. 1998), and a type-independent ``sandwich-model''
(Matthews et al. 1999).  We followed the method of Tully et al. (1998)
and calculated the internal extinction of a face-on galaxy in the
$I$-band ($\gamma_I$), using the equation:

\begin{equation}
\gamma_{I}=0.92+1.63(\log(2V_{rot})-2.5)
\end{equation}

\noindent where $2V_{rot}$ is set to the inclination corrected $W_{20}$
derived in Garcia-Appadoo et al. (2009).  Several of the galaxies do
not have well-measured axis ratios and thus do not have well measured
inclinations, making the appropriate value of $V_{rot}$ uncertain.  An
inclination of 60 degrees (the average inclination in a randomly
aligned sample) was assigned to all of these galaxies.  The total extinction correction for the inclined galaxy was then calculated from $\gamma_I$ using:

\begin{equation}
A_I=\gamma_{I}\log\left(\frac{1}{b/a}\right)
\end{equation}

\noindent where $b/a$ is the axis ratio of the galaxy calculated from the S\'{e}rsic fits described Section 4.3.  

We converted the $I$ band extinction to the SDSS bands using the
relations in Schlegel et al. (1998).  The extinction values relative
to $I$-band for $u$, $g$, $r$, $i$, and $z$ are: 2.66, 1.95, 1.42,
1.07, and 0.763 respectively.  Many of the calculated extinction
values for galaxies with uncertain axis ratios were negative and all
of them had values below 0.09 magnitudes in the $I$-band.  These were converted
to 0 before being applied to the galaxy photometry.  The extinction
values computed for galaxies with uncertain inclinations are therefore
unlikely to have produced significant errors in the final photometry.
These galaxies are primarily very blue, low surface brightness systems
with low metallicities, and are unlikely to have large dust components. All of the internal extinction values are included in Table \ref{table:photcor}.


\subsubsection{$K$-Corrections}

Although this sample of galaxies is nearby, $K$-corrections are small
but important for precise photometry of the more distant galaxies in
the sample.  We used Blanton et al.'s (2003b) {\tt kcorrect\_v3.2} to
$K$-correct all of the galaxies in the sample to $z$=0.  These
corrections are included in Table \ref{table:photcor}.  The median
and the maximum $r$-band k-corrections are 0.008 and 0.031
respectively.  For the $u$-band, the median and maximum k-corrections
are more important at 0.012 and 0.070 respectively. 



\subsection{Other Measured Petrosian Properties}

In addition to providing a measurement of flux, Petrosian quantities
can be used to calculate robust measurements of size and surface
brightness.  In this section, we describe the method used to derive
these quantities.

Once a Petrosian flux was measured (see above), we computed the 50\%
(R50) and 90\% (R90) radii for both the elliptical apertures and
circular apertures.  These values correspond to the {\tt PetroR50} and
{\tt PetroR90} parameters in the SDSS {\tt tsObj} files.  They were
converted to arcseconds using the SDSS pixel scale of
0.396$^{\prime\prime}$/pixel.  The best-fit R50 and R90 values and
their uncertainties are reported in Table \ref{table:paper}. For an
exponential disk galaxy, the R50 and R90 values correspond to 1.668
and 3.816 times the scale length respectively.  For a de Vaucouleurs
model the R50 and R90 correspond to 0.7124 and 2.387 times the
effective half-light radius ($r_e$).

The Petrosian flux and sizes can be used to compute surface brighteness for each galaxy.  The Petrosian surface brightness is the average surface brightness within R50, defined by taking half of the Petrosian flux and dividing by the elliptical
aperture area with a semi-major axis equal to the elliptical R50. For
galaxies with well defined elliptical apertures, this is the most
reliable measure of the average surface brightness of the galaxy as it
appears in the image. For a face-on exponential disk, the Petrosian surface
brightness is 1.118 times the central surface brightness in
mag/$\Box^{\prime\prime}$.

\section{Photometric Sample}
\subsection{Derived Quantities}

With excellent five-band photometry, it is possible to explore the
photometric properties of this HI-selected sample with unprecedented
uniformity.  In this section we derive several quantities from the
photometric data, give an overview of the sample properties. and
highlight some empirical results from the photometry.

Using the distances derived from the Virgo infall model described
above, we computed physical sizes and absolute magnitudes for every
galaxy, given in Table \ref{table:derive}.  There are uncertainties in the flow models used to calculate the Virgo infall velocities.  When we computed the distances without the Virgo infall corrections, we get values that differ by at most 6\% (average deviation of 2\%).  When we compared our Virgo infall derived distances to those computed in relation to the Cosmic Microwave Background, we found deviations of $\sim$13\%.  We chose a conservative error uncertainty of 13\% for our distance estimates and propagated this uncertainty when calculating absolute magnitudes, stellar masses and physical sizes.

Absolute magnitudes were calculated using the standard distance
modulus formula including all photometric corrections (foreground
extinction, internal extinction, $K$-corrections).  Luminosities can
be computed from the tabulated values of absolute magnitude using the
solar absolute $ugriz$ magnitudes (6.41, 5.15, 4.67, 4.56, and 4.53
magnitudes respectively), provided by Bell et al. (2003).  We also
tabulated the physical R50 and R90 size (kpc) of each galaxy in the
$r$-band.

To aid in Tully-Fisher and other dynamical studies, we included the
inclination and turbulence corrected circular velocity in Table
\ref{table:derive}.  We correct for inclination and turbulence effects
in the HI line widths using the equation:

\begin{equation}
W_{20,c}=\frac{W_{20}-W_{20,t}}{2}\left(\frac{1-(b/a)^2}{1-0.19^2}\right)^{-1/2}
\end{equation}

\noindent where $b/a$ is the optically derived axis ratio, $W_{20,t}$
is the turbulence correction and 0.19 is typical intrinsic axis ratio
for spiral galaxies (Pizagno et al. 2005).  For $W_{20,t}$, we use the
empirically derived value of 8 km\ s$^{-1}$ from the late-type galaxy
study of Begum et al. (2006).  Because of the large uncertainties in
several of the derived inclinations (galaxies with circular aperture
photometry), not all of the velocities could be properly corrected.

To compute stellar masses, we used the method of Bell et al. (2003) to
calculate the stellar mass-to-light ratio  in the $i$-band.
Specifically we used the relation:
\begin{equation}
\log(M_{\ast}/L_{i})=-0.222+0.864(g-r)+\log(0.71)
\end{equation}

\noindent where the factor of 0.71 comes from the conversion between
Bell et al.'s (2003)``diet Salpeter'' and a Kroupa IMF (Pizagno et
al. 2005). The $g-r$ color is used to compute the stellar
mass-to-light ratio because it is robust against large changes in
color when emission lines dominate the galaxy's spectral energy
distribution (West et al. 2009).  We used the $i$-band magnitude to
compute the stellar mass for the same reason; the $i$-band has very
few emission lines and is a good indicator of the underlying stellar
population.

Assuming an $i$-band solar absolute magnitude of 4.56 (Bell et
al. 2003), we derived the stellar mass using:
\begin{equation}
M_{\ast}=M_{\ast}/L_{i}\times10^{(-(M_{i}-4.56)/2.51)}.
\end{equation}

The stellar masses are given in Table \ref{table:derive}.

\subsection{Sample Properties}

\begin{figure}
\centering
\plotone{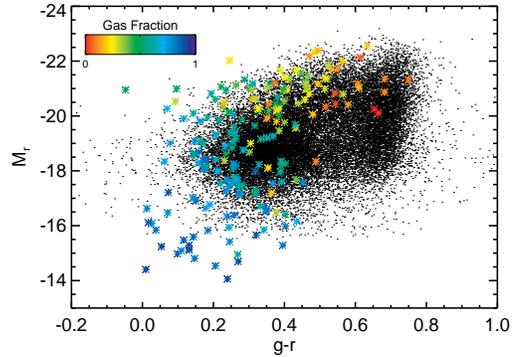}
\caption{Absolute $r$-band magnitude as a function of $g-r$ for the HI-selected galaxies (color asterisks) and the SDSS DR4 main galaxy sample (black dots). ES/SDSS galaxies have been color coded according to their gas fractions. The HI-selection identifies a bluer and less luminous sample of galaxies than those used in typical SDSS studies.}
\label{migr}
\end{figure}

With the photometry and other photometric quantities complete, we
briefly examined the global photometric properties of the sample.  The
ES/SDSS sample selects galaxies that are bluer, less luminous, less
massive, and undergoing higher rates of star formation than those in
the SDSS main sample.  As we discussed above, a large fraction of
these differences are based on the spectral targeting (which does not
select LSB galaxies), the photometric pipeline problems (which make
nearby galaxies in SDSS difficult to analyze), and the fact that we
sampled a very different volume than typical SDSS studies (see Fig.
\ref{zdist}).  Despite, the heavy selection effects, it is important
to demonstrate how an HI-selected sample produces a drastically
different sample than a typical SDSS galaxy study. These differences
can be seen in Fig. \ref{migr}, the $M_r$ vs. $g-r$ color-magnitude
diagram for the HI-selected galaxies (color asterisks) and the DR4
SDSS main galaxy sample (black dots).  All of the photometric corrections have been applied except for the internal extinction, which requires information about the rotation velocity (a quantity that is not available for the DR4 galaxies).  The ES/SDSS data have been
color coded according to their gas fractions.\footnote{Gas fraction is
  defined as the mass in HI (corrected for heavier elements) divided
  by the sum of the mass in HI and stars
  ($f_{gas}=1.4{\rm{M}}_{\rm{HI}}/(1.4{\rm{M}}_{\rm{HI}}+{\rm{M}}_{\ast})$).}
While the bimodal distribution of DR4 galaxies, can be clearly seen in
Fig. \ref{migr}, the ES/SDSS sample dominates the blue distribution.
The galaxies with the highest gas fractions occupy the bluest,
lowest-luminosity region of color-magnitude space. These trends
between color, luminosity and gas content are investigated in other
studies (Garcia-Appadoo et al. 2009; West et al. 2009).  Because of
the way that SDSS galaxies are selected for both spectroscopy and
scientific analysis, most SDSS studies do not include the bluest,
faintest galaxies that are easily identified in an HI-selected sample.

\begin{figure}
\centering
\plotone{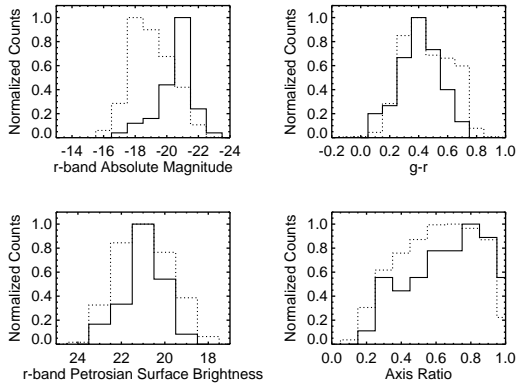}
\caption{Normalized distributions of $r$-band absolute magnitude (top left), $g-r$ color (top right), $r$-band Petrosian surface brightness (bottom left) and axis ratio (bottom right) for the ES/SDSS (solid) and SDSS DR4 (dotted) galaxies that fall within the overlap region of the two samples (40 Mpc $<$ $D$ $<$ 140 Mpc). While the surface brightness distributions are similar, the ES/SDSS galaxies are bluer, intrinsically brighter and more likely to be face-on than the DR4 galaxies in the same volume.}
\label{compdr4}
\end{figure}

To eliminate some of the selection effects due to sampling different volumes, we limited our comparison between the DR4 and ES/SDSS galaxies to those that fall within the overlap region of the two samples (40 Mpc $<$ $D$ $<$ 140 Mpc).  There are 51 ES/SDSS and 18,379 DR4 galaxies in this volume.  Figure \ref{compdr4} shows the normalized distributions for four measured or derived properties of the ES/SDSS (solid) and SDSS DR4 (dotted) galaxies.  We calculated the mean values for each distribution as well as computed the Kolmogorov-Smirnov (KS) statistic to determine how likely the two histograms were drawn from the same distribution.  The $r$-band Petrosian surface brightness distributions (bottom left) of the two populations are very similar, with both the ES/SDSS and DR4 samples having mean values of 20.9 mag/arcsec$^2$ and a KS probability of 0.19 (19\% chance of being pulled from the same distribution).  The ES/SDSS galaxies are both bluer (average $g-r$ values of 0.40 versus 0.47; top right) and intrinsically brighter ($M_r$ values of -20.6 versus -19.0; top left) than the DR4 SDSS main galaxies in the same volume. Both the properties have KS probabilities that give them a $<$ 3\% chance of the ES/SDSS and DR4 galaxies being pulled from the same distribution.  Some of the discrepancy between the distributions is due to a significant population of dwarf elliptical galaxies in the DR4 sample that have high surface brightnesses (and thus are targeted for spectroscopy) and have small or non-existent quantities of gas (and are omitted from an HI-selected sample).  In addition, the bluer $g-r$ colors in the ES/SDSS galaxies may be indicative of a recent burst of star formation in galaxies with a reservoir of gas (West et al. 2009).

The axis ratio distribution demonstrates a slight bias towards face-on galaxies in the ES/SDSS sample.  The mean axis ratios are 0.67 and 0.61 for the ES/SDSS and DR4 samples respectively.  There is a 7\% chance that the histograms are pulled from the same distribution.  While the SDSS selection should be almost independent of axis ratio (and therefore be fairly representative of the population), HI selected samples should have a slight bias towards face-on systems (the same amount of HI flux will be distributed into fewer channels in a face-on galaxy, resulting in brighter peak flux and easier detection).  Figure \ref{compdr4} highlights an intrinsic bias in HI-selected surveys that will be important to consider for the next-generation HI-surveys like ALFALFA (Giovanelli et al. 2005), which will include tens of thousands of HI-selected sources.

\section{Summary}

Using our customized software for analyzing large galaxies in the SDSS, we
have created an HI-selected catalog using data from the ES and
SDSS surveys.  The data are uniformly sampled and have well
characterized uncertainties and limitations. Our software addresses the deblending and sky subtraction issues that make SDSS catalog-level photometry of angularly large galaxies in SDSS unreliable.  We specifically discussed the problems with deblending and sky subtraction and quantified their effect on the ES/SDSS sample.  In addition, we derived a hyper-plane that relates the amount of flux removed by the sky subtraction algorithm to the size, magnitude and axis ratio of the galaxy.

The ES/SDSS galaxies span a large range of surface brightnesses,
colors and stellar masses and as expected, the optical properties are
different from those of most SDSS galaxy samples (mostly due to selection effects), as well as an optically selected sample compared at the sample volume.  Specifically, the HI selection identifies galaxies with lower surface brightness, smaller absolute magnitudes, bluer colors and smaller stellar masses than those used in typical SDSS studies.

HI-selected galaxies (compared to optically selected galaxies in the same volume) are bluer and brighter, the latter due to the abundance of dwarf elliptical galaxies in optically selected samples.  The ES/SDSS sample also shows a slight bias towards face-on systems, a symptom of the HI-selection.

Other papers incorporate the ES HI data and investigate how the SDSS
photometric trends relate to the galaxy gas content (Garcia-Appadoo et
al. 2009) and attempt to understand the colors of the galaxies (West
et al. 2009).  Future work will examine the dynamics of the ES/SDSS
sample and investigate the Tully Fisher relation and its scatter.

\section{Acknowledgments} 

The authors gratefully acknowledge the financial support of the
Astronaut Scholarship Foundation, ADVANCE, ARCS, the Royalty Research
Fund, NSF grant 0540567, the Theodore Jacobsen Fund.  We also would
like to extend our gratitude to the anonymous referee who provided a
substantial amount of feedback, which was extremely useful to the
completion of this paper. AAW thanks David Hogg, Mike Blanton, Marla
Geha, Beth Willman, Anil Seth, Adam Burgasser, Kevin Covey, John
Bochanski and Maritza Tavarez-Brown for their helpful comments,
suggestions, financial support and gentle prodding.

Funding for the SDSS and SDSS-II has been provided by the Alfred
P. Sloan Foundation, the Participating Institutions, the National
Science Foundation, the U.S. Department of Energy, the National
Aeronautics and Space Administration, the Japanese Monbukagakusho, the
Max Planck Society, and the Higher Education Funding Council for
England. The SDSS Web Site is http://www.sdss.org/.

The SDSS is managed by the Astrophysical Research Consortium for the
Participating Institutions. The Participating Institutions are the
American Museum of Natural History, Astrophysical Institute Potsdam,
University of Basel, Cambridge University, Case Western Reserve
University, University of Chicago, Drexel University, Fermilab, the
Institute for Advanced Study, the Japan Participation Group, Johns
Hopkins University, the Joint Institute for Nuclear Astrophysics, the
Kavli Institute for Particle Astrophysics and Cosmology, the Korean
Scientist Group, the Chinese Academy of Sciences (LAMOST), Los Alamos
National Laboratory, the Max-Planck-Institute for Astronomy (MPIA),
the Max-Planck-Institute for Astrophysics (MPA), New Mexico State
University, Ohio State University, University of Pittsburgh,
University of Portsmouth, Princeton University, the United States
Naval Observatory, and the University of Washington.



\begin{thebibliography}{66}
\expandafter\ifx\csname natexlab\endcsname\relax\def\natexlab#1{#1}\fi

\bibitem[{{Abazajian} {et~al.}(2009)}]{DR7}
{Abazajian}, K.~N., {et~al.} 2009, \apjs, 182, 543

\bibitem[{{Abazajian} {et~al.}(2004)}]{2004AJ....128..502A}
{Abazajian}, K., {et~al.} 2004, \aj, 128, 502

\bibitem[{{Adelman-McCarthy} {et~al.}(2008)}]{DR6}
{Adelman-McCarthy}, J.~K. {et~al.} 2008, \apjs, 175, 297

\bibitem[{{Ag{\"u}eros} {et~al.}(2005)}]{2005AJ....130.1022A}
{Ag{\"u}eros}, M.~A., {et~al.} 2005, \aj, 130, 1022

\bibitem[{{Bailin} \& {Harris}(2008)}]{Bailin08}
{Bailin}, J., \& {Harris}, W.~E. 2008, \apj, 681, 225

\bibitem[{{Baldry} {et~al.}(2005){Baldry}, {Glazebrook}, {Budav{\'a}ri},
  {Eisenstein}, {Annis}, {Bahcall}, {Blanton}, {Brinkmann}, {Csabai},
  {Heckman}, {Lin}, {Loveday}, {Nichol}, \& {Schneider}}]{2005MNRAS.358..441B}
{Baldry}, I.~K., {Glazebrook}, K., {Budav{\'a}ri}, T., {Eisenstein}, D.~J.,
  {Annis}, J., {Bahcall}, N.~A., {Blanton}, M.~R., {Brinkmann}, J., {Csabai},
  I., {Heckman}, T.~M., {Lin}, H., {Loveday}, J., {Nichol}, R.~C., \&
  {Schneider}, D.~P. 2005, \mnras, 358, 441

\bibitem[{{Barnes} {et~al.}(2001)}]{2001MNRAS.322..486B}
{Barnes}, D.~G., {et~al.} 2001, \mnras, 322, 486

\bibitem[{{Begum} {et~al.}(2006){Begum}, {Chengalur}, {Karachentsev}, {Kaisin},
  \& {Sharina}}]{2006MNRAS.365.1220B}
{Begum}, A., {Chengalur}, J.~N., {Karachentsev}, I.~D., {Kaisin}, S.~S., \&
  {Sharina}, M.~E. 2006, \mnras, 365, 1220

\bibitem[{{Bell} {et~al.}(2003){Bell}, {McIntosh}, {Katz}, \&
  {Weinberg}}]{2003ApJS..149..289B}
{Bell}, E.~F., {McIntosh}, D.~H., {Katz}, N., \& {Weinberg}, M.~D. 2003, \apjs,
  149, 289

\bibitem[{{Bernardi}(2007)}]{Bernardi07}
{Bernardi}, M. 2007, \aj, 133, 1954

\bibitem[{{Bertin} \& {Arnouts}(1996)}]{1996A&AS..117..393B}
{Bertin}, E., \& {Arnouts}, S. 1996, \aaps, 117, 393


\bibitem[{{Blanton} {et~al.}(2005)}]{Blanton08}
{Blanton}, M.~R., {Geha}, M., \& {West}, A.~A. 2008, \apj, 682,861

\bibitem[{{Blanton} {et~al.}(2005){Blanton}, {Schlegel}, {Strauss},
  {Brinkmann}, {Finkbeiner}, {Fukugita}, {Gunn}, {Hogg}, {Ivezi{\'c}}, {Knapp},
  {Lupton}, {Munn}, {Schneider}, {Tegmark}, \& {Zehavi}}]{2005AJ....129.2562B}
{Blanton}, M.~R., {et~al.}  2005, \aj, 129, 2562

\bibitem[{{Blanton} {et~al.}(2003{\natexlab{a}}){Blanton}, {Brinkmann},
  {Csabai}, {Doi}, {Eisenstein}, {Fukugita}, {Gunn}, {Hogg}, \&
  {Schlegel}}]{2003AJ....125.2348B}
{Blanton}, M.~R., {Brinkmann}, J., {Csabai}, I., {Doi}, M., {Eisenstein}, D.,
  {Fukugita}, M., {Gunn}, J.~E., {Hogg}, D.~W., \& {Schlegel}, D.~J.
  2003{\natexlab{a}}, \aj, 125, 2348


\bibitem[{{Blanton} {et~al.}(2003{\natexlab{b}})}]{2003ApJ...594..186B}
---. 2003{\natexlab{b}}, \apj, 594, 186

\bibitem[{{Blanton} {et~al.}(2001)}]{2001AJ....121.2358B}
{Blanton}, M.~R., {et~al.} 2001, \aj, 121, 2358


\bibitem[{{Burkholder} {et~al.}(2001){Burkholder}, {Impey}, \&
  {Sprayberry}}]{2001AJ....122.2318B}
{Burkholder}, V., {Impey}, C., \& {Sprayberry}, D. 2001, \aj, 122, 2318

\bibitem[{{Courteau}(1996)}]{1996ApJS..103..363C}
{Courteau}, S. 1996, \apjs, 103, 363

\bibitem[{{Covey} {et~al.}(2007)}]{2007AJ....134.2398C}
{Covey}, K.~R., {et~al.} 2007, \aj, 134, 2398

\bibitem[{{Disney} {et~al.}(2008)}]{Disney08}
{Disney}, M.~J., {Romano}, J.~D., {Garcia-Appadoo}, D.~A., {West}, A.~A., {Dalcanton}, J.~J., \& {Cortese}, L. 2008, \nat, 455, 1082

\bibitem[{{Disney}(1976)}]{1976Natur.263..573D}
{Disney}, M.~J. 1976, \nat, 263, 573

\bibitem[{{Doyle} {et~al.}(2005)}]{2005MNRAS.361...34D}
{Doyle}, M.~T., {et~al.} 2005, \mnras, 361, 34

\bibitem[{{Fisher} \& {Tully}(1981)}]{1981ApJS...47..139F}
{Fisher}, J.~R., \& {Tully}, R.~B. 1981, \apjs, 47, 139

\bibitem[{{Fukugita} {et~al.}(1996){Fukugita}, {Ichikawa}, {Gunn}, {Doi},
  {Shimasaku}, \& {Schneider}}]{1996AJ....111.1748F}
{Fukugita}, M., {Ichikawa}, T., {Gunn}, J.~E., {Doi}, M., {Shimasaku}, K., \&
  {Schneider}, D.~P. 1996, \aj, 111, 1748

\bibitem[{{Garcia-Appadoo} {et~al.}(2009)}]{Garcia09}
{Garcia-Appadoo}, D.~A., {West}, A.~A., {Dalcanton}, J.~J., {Cortese}, L., \& {Disney}, M.~J. 2009, /mnras, 394, 340

\bibitem[{{Gavazzi} \& {Boselli}(1996)}]{1996ApL&C..35....1G}
{Gavazzi}, G., \& {Boselli}, A. 1996, Astrophysical Letters Communications, 35,
  1

\bibitem[{{Geha} {et~al.}(2006){Geha}, {Blanton}, {Masjedi}, \&
  {West}}]{2006ApJ}
{Geha}, M., {Blanton}, M.~R., {Masjedi}, M., \& {West}, A.~A. 2006, \apj, 999

\bibitem[{{Giovanelli} {et~al.}(2005)}]{ALFALFA}
{Giovanelli}, R., {et~al.} 2005, \aj, 130, 2598

\bibitem[{{Gunn} {et~al.}(1998)}]{1998AJ....116.3040G}
{Gunn}, J.~E., {et~al.} 1998, \aj, 116, 3040

\bibitem[{{Gunn} {et~al.}(2006)}]{2006AJ....131.2332G}
---. 2006, \aj, 131, 2332

\bibitem[{{Hambly} {et~al.}(2001{\natexlab{a}}){Hambly}, {Davenhall}, {Irwin},
  \& {MacGillivray}}]{2001MNRAS.326.1315H}
{Hambly}, N.~C., {Davenhall}, A.~C., {Irwin}, M.~J., \& {MacGillivray}, H.~T.
  2001{\natexlab{a}}, \mnras, 326, 1315

\bibitem[{{Hambly} {et~al.}(2001{\natexlab{b}}){Hambly}, {Irwin}, \&
  {MacGillivray}}]{2001MNRAS.326.1295H}
{Hambly}, N.~C., {Irwin}, M.~J., \& {MacGillivray}, H.~T. 2001{\natexlab{b}},
  \mnras, 326, 1295

\bibitem[{{Hambly} {et~al.}(2001{\natexlab{c}})}]{2001MNRAS.326.1279H}
{Hambly}, N.~C., {et~al.} 2001{\natexlab{c}}, \mnras, 326, 1279

\bibitem[{{Haynes} {et~al.}(1999){Haynes}, {Giovanelli}, {Chamaraux}, {da
  Costa}, {Freudling}, {Salzer}, \& {Wegner}}]{1999AJ....117.2039H}
{Haynes}, M.~P., {Giovanelli}, R., {Chamaraux}, P., {da Costa}, L.~N.,
  {Freudling}, W., {Salzer}, J.~J., \& {Wegner}, G. 1999, \aj, 117, 2039

\bibitem[{{Helmboldt} {et~al.}(2004){Helmboldt}, {Walterbos}, {Bothun},
  {O'Neil}, \& {de Blok}}]{2004ApJ...613..914H}
{Helmboldt}, J.~F., {Walterbos}, R.~A.~M., {Bothun}, G.~D., {O'Neil}, K., \&
  {de Blok}, W.~J.~G. 2004, \apj, 613, 914

\bibitem[{{Henning}(1992)}]{1992ApJS...78..365H}
{Henning}, P.~A. 1992, \apjs, 78, 365

\bibitem[{{Henning}(1995)}]{1995ApJ...450..578H}
---. 1995, \apj, 450, 578

\bibitem[{{Hogg} {et~al.}(2001){Hogg}, {Finkbeiner}, {Schlegel}, \&
  {Gunn}}]{2001AJ....122.2129H}
{Hogg}, D.~W., {Finkbeiner}, D.~P., {Schlegel}, D.~J., \& {Gunn}, J.~E. 2001,
  \aj, 122, 2129

\bibitem[{{Hopkins} {et~al.}(2003)}]{2003ApJ...599..971H}
{Hopkins}, A.~M., {et~al.} 2003, \apj, 599, 971

\bibitem[{{Iglesias-P{\' a}ramo} {et~al.}(2003){Iglesias-P{\' a}ramo}, {van
  Driel}, {Duc}, {Papaderos}, {V{\'{\i}}lchez}, {Cayatte}, {Balkowski},
  {O'Neil}, {Dickey}, {Hern{\' a}ndez}, \& {Thuan}}]{2003A&A...406..453I}
{Iglesias-P{\' a}ramo}, J., {van Driel}, W., {Duc}, P.-A., {Papaderos}, P.,
  {V{\'{\i}}lchez}, J.~M., {Cayatte}, V., {Balkowski}, C., {O'Neil}, K.,
  {Dickey}, J., {Hern{\' a}ndez}, H., \& {Thuan}, T.~X. 2003, \aap, 406, 453

\bibitem[{{Ivezi{\'c}} {et~al.}(2004)}]{2004AN....325..583I}
{Ivezi{\'c}}, {\v Z}., {et~al.} 2004, Astronomische Nachrichten, 325, 583

\bibitem[{{Jarrett} {et~al.}(2000){Jarrett}, {Chester}, {Cutri}, {Schneider},
  {Skrutskie}, \& {Huchra}}]{2000AJ....119.2498J}
{Jarrett}, T.~H., {Chester}, T., {Cutri}, R., {Schneider}, S., {Skrutskie}, M.,
  \& {Huchra}, J.~P. 2000, \aj, 119, 2498

\bibitem[{{Karachentsev} {et~al.}(2004){Karachentsev}, {Karachentseva},
  {Huchtmeier}, \& {Makarov}}]{2004AJ....127.2031K}
{Karachentsev}, I.~D., {Karachentseva}, V.~E., {Huchtmeier}, W.~K., \&
  {Makarov}, D.~I. 2004, \aj, 127, 2031

\bibitem[{{Kennicutt} {et~al.}(1994){Kennicutt}, {Tamblyn}, \&
  {Congdon}}]{1994ApJ...435...22K}
{Kennicutt}, R.~C., {Tamblyn}, P., \& {Congdon}, C.~E. 1994, \apj, 435, 22

\bibitem[{{Kron}(1980)}]{1980ApJS...43..305K}
{Kron}, R.~G. 1980, \apjs, 43, 305

\bibitem[{{Lauer} {et~al.}(2007)}]{Lauer07}
{Lauer}, T.~R., {et~al.} 2007, \apj, 662, 808

\bibitem[{{Lupton} {et~al.}(1999){Lupton}, {Gunn}, \&
  {Szalay}}]{1999AJ....118.1406L}
{Lupton}, R.~H., {Gunn}, J.~E., \& {Szalay}, A.~S. 1999, \aj, 118, 1406

\bibitem[{{Lupton} {et~al.}(2002){Lupton}, {Ivezic}, {Gunn}, {Knapp},
  {Strauss}, \& {Yasuda}}]{2002SPIE.4836..350L}
{Lupton}, R.~H., {Ivezic}, Z., {Gunn}, J.~E., {Knapp}, G., {Strauss}, M.~A., \&
  {Yasuda}, N. 2002, in Survey and Other Telescope Technologies and
  Discoveries. Edited by Tyson, J. Anthony; Wolff, Sidney. Proceedings of the
  SPIE, Volume 4836, pp. 350-356 (2002)., ed. J.~A. {Tyson} \& S.~{Wolff},
  350--356

\bibitem[{{MacArthur} {et~al.}(2003){MacArthur}, {Courteau}, \&
  {Holtzman}}]{2003ApJ...582..689M}
{MacArthur}, L.~A., {Courteau}, S., \& {Holtzman}, J.~A. 2003, \apj, 582, 689

\bibitem[{{Mandelbaum} {et~al.}(2005)}]{Mandelbaum05}
{Mandelbaum}, R., {Hirata}, C.~M., {Seljak}, U., {Guzik}, J., {Padmanabhan}, N., {Blake}, C., {Blanton}, M.~R., {Lupton}, R., \& {Brinkmann}, J. 2005, \mnras, 361, 1287

\bibitem[{{Masjedi} {et~al.}(2006)}]{Masjedi06}
{Masjedi}, M., {et~al.} 2006, \apj, 644, 54

\bibitem[{{Matthews} {et~al.}(1999){Matthews}, {Gallagher}, \& {van
  Driel}}]{1999AJ....118.2751M}
{Matthews}, L.~D., {Gallagher}, J.~S., \& {van Driel}, W. 1999, \aj, 118, 2751

\bibitem[{{McGaugh} \& {de Blok}(1997)}]{1997ApJ...481..689M}
{McGaugh}, S.~S., \& {de Blok}, W.~J.~G. 1997, \apj, 481, 689

\bibitem[{{Meyer} {et~al.}(2004)}]{2004MNRAS.350.1195M}
{Meyer}, M.~J., {et~al.} 2004, \mnras, 350, 1195

\bibitem[{{Obri{\'c}} {et~al.}(2006)}]{2006MNRAS}
{Obri{\'c}}, M., {et~al.} 2006, MNRAS, 999

\bibitem[{{Petrosian}(1976)}]{1976ApJ...209L...1P}
{Petrosian}, V. 1976, \apjl, 209, L1

\bibitem[{{Pier} {et~al.}(2003){Pier}, {Munn}, {Hindsley}, {Hennessy}, {Kent},
  {Lupton}, \& {Ivezi{\' c}}}]{2003AJ....125.1559P}
{Pier}, J.~R., {Munn}, J.~A., {Hindsley}, R.~B., {Hennessy}, G.~S., {Kent},
  S.~M., {Lupton}, R.~H., \& {Ivezi{\' c}}, {\v Z}. 2003, \aj, 125, 1559

\bibitem[{{Pizagno} {et~al.}(2005){Pizagno}, {Prada}, {Weinberg}, {Rix},
  {Harbeck}, {Grebel}, {Bell}, {Brinkmann}, {Holtzman}, \&
  {West}}]{2005ApJ...633..844P}
{Pizagno}, J., {Prada}, F., {Weinberg}, D.~H., {Rix}, H.-W., {Harbeck}, D.,
  {Grebel}, E.~K., {Bell}, E.~F., {Brinkmann}, J., {Holtzman}, J., \& {West},
  A. 2005, \apj, 633, 844

\bibitem[{{Roberts}(1963)}]{1963ARA&A...1..149R}
{Roberts}, M.~S. 1963, \araa, 1, 149

\bibitem[{{Rosenberg} \& {Schneider}(2000)}]{2000ApJS..130..177R}
{Rosenberg}, J.~L., \& {Schneider}, S.~E. 2000, \apjs, 130, 177

\bibitem[{{Rosenberg} \& {Schneider}(2002)}]{2002ApJ...567..247R}
---. 2002, \apj, 567, 247

\bibitem[{{Rosenberg} {et~al.}(2005){Rosenberg}, {Schneider}, \&
  {Posson-Brown}}]{2005AJ....129.1311R}
{Rosenberg}, J.~L., {Schneider}, S.~E., \& {Posson-Brown}, J. 2005, \aj, 129,
  1311

\bibitem[{{S{\' e}rsic}(1968)}]{1968BAAA...13Q..20S}
{S{\' e}rsic}, J.~L. 1968, Boletin de la Asociacion Argentina de Astronomia La
  Plata Argentina, 13, 20

\bibitem[{{Salim} {et~al.}(2005)}]{2005ApJ...619L..39S}
{Salim}, S., {et~al.} 2005, \apjl, 619, L39

\bibitem[{Serra} {et~al.}]{Serra07}
{Serra}, P., {Trager}, S.~C., {van der Hulst}, J.~M.,
	{Oosterloo}, T.~A., {Morganti}, R., {van Gorkom}, J.~H. \& 
	{Sadler}, E.~M. 2007, New Astronomy Reviews, 51, 3


\bibitem[{{Schlegel} {et~al.}(1998){Schlegel}, {Finkbeiner}, \&
  {Davis}}]{1998ApJ...500..525S}
{Schlegel}, D.~J., {Finkbeiner}, D.~P., \& {Davis}, M. 1998, \apj, 500, 525

\bibitem[{{Scodeggio} \& {Gavazzi}(1993)}]{1993ApJ...409..110S}
{Scodeggio}, M., \& {Gavazzi}, G. 1993, \apj, 409, 110

\bibitem[{{Skrutskie} {et~al.}(2006)}]{2006AJ....131.1163S}
{Skrutskie}, M.~F., {et~al.} 2006, \aj, 131, 1163

\bibitem[{{Smith} {et~al.}(2002)}]{2002AJ....123.2121S}
{Smith}, J.~A., {et~al.} 2002, \aj, 123, 2121

\bibitem[{{Smol{\v c}i{\'c}} {et~al.}(2006)}]{2006MNRAS.371..121S}
{Smol{\v c}i{\'c}}, V., {et~al.} 2006, \mnras, 371, 121

\bibitem[{{Spitzak} \& {Schneider}(1998)}]{1998ApJS..119..159S}
{Spitzak}, J.~G., \& {Schneider}, S.~E. 1998, \apjs, 119, 159

\bibitem[{{Staveley-Smith} {et~al.}(1996)}]{1996PASA...13..243S}
{Staveley-Smith}, L., {et~al.} 1996, PASA, 13, 243

\bibitem[{{Stoughton} {et~al.}(2002)}]{2002AJ....123..485S}
{Stoughton}, C., {et~al.} 2002, \aj, 123, 485

\bibitem[{{Strauss} {et~al.}(2002)}]{2002AJ....124.1810S}
{Strauss}, M.~A., {et~al.} 2002, \aj, 124, 1810

\bibitem[{{Swaters} {et~al.}(2002){Swaters}, {van Albada}, {van der Hulst}, \&
  {Sancisi}}]{2002A&A...390..829S}
{Swaters}, R.~A., {van Albada}, T.~S., {van der Hulst}, J.~M., \& {Sancisi}, R.
  2002, \aap, 390, 829

\bibitem[{{Terry} {et~al.}(2002){Terry}, {Paturel}, \&
  {Ekholm}}]{2002A&A...393...57T}
{Terry}, J.~N., {Paturel}, G., \& {Ekholm}, T. 2002, \aap, 393, 57

\bibitem[{{Theureau} {et~al.}(1998){Theureau}, {Rauzy}, {Bottinelli}, \&
  {Gouguenheim}}]{1998A&A...340...21T}
{Theureau}, G., {Rauzy}, S., {Bottinelli}, L., \& {Gouguenheim}, L. 1998, \aap,
  340, 21

\bibitem[{{Tucker} {et~al.}(2006)}]{2006AN....327..821T} {Tucker}, D.~L., {et~al.} 2006, Astronomische Nachrichten, 327, 821

\bibitem[{{Tully} {et~al.}(1998){Tully}, {Pierce}, {Huang}, {Saunders},
  {Verheijen}, \& {Witchalls}}]{1998AJ....115.2264T}
{Tully}, R.~B., {Pierce}, M.~J., {Huang}, J., {Saunders}, W., {Verheijen},
  M.~A.~W., \& {Witchalls}, P.~L. 1998, \aj, 115, 2264

\bibitem[{{Walter} {et~al.}(2008)}]{Walter08}
{Walter}, F., {Brinks}, E., {de Blok}, W.~J.~G., {Bigiel}, F., {Kennicutt}, R.~C., {Thornley}, M.~D., \& {Leroy}, A. 2008, \aj, 136, 2563

\bibitem[{{West} {et~al.}(2009)}]{West09}
{West}, A.~A., {Garcia-Appadoo}, D.~A., {Dalcanton}, J.~J., {Disney}, M.~J., {Rockosi}, C.~M., \& {Ivezi{\'c}}, {\v Z}. 2009, \aj, 138, 796

\bibitem[{{Wong} {et~al.}(2009)}]{Wong09}
{Wong}, O.~I., {Webster}, R., {Kilborn}, V., {Waugh}, M., \&
	{Staveley-Smith}, L. 2009, \mnras, in press

\bibitem[{{Wong} {et~al.}(2006)}]{2006MNRAS.371.1855W}
{Wong}, O.~I., {et~al.} 2006, \mnras, 371, 1855

\bibitem[{{Yasuda} {et~al.}(2001)}]{2001AJ....122.1104Y}
{Yasuda}, N., {et~al.} 2001, \aj, 122, 1104

\bibitem[{{York} {et~al.}(2000)}]{2000AJ....120.1579Y}
{York}, D.~G., {et~al.} 2000, \aj, 120, 1579

\bibitem[{Zhang} {et~al.}(2009)]{zhang09}
{Zhang}, W., {Li}, C., {Kauffmann}, G., {Zou}, H., {Catinella}, B., {Shen}, S., {Guo}, Q., \& {Chang}, R. 2009, \mnras, in press
	

\bibitem[{{Zwaan} {et~al.}(1997){Zwaan}, {Briggs}, {Sprayberry}, \&
  {Sorar}}]{1997ApJ...490..173Z}
{Zwaan}, M.~A., {Briggs}, F.~H., {Sprayberry}, D., \& {Sorar}, E. 1997, \apj,
  490, 173

\bibitem[{{Zwaan} {et~al.}(2004)}]{2004MNRAS.350.1210Z}
{Zwaan}, M.~A., {et~al.} 2004, \mnras, 350, 1210

\end{thebibliography}

\clearpage



\clearpage
\begin{landscape}

\begin{deluxetable}{lccccccclc}
\tablewidth{0pt}
\tablecolumns{10} 
\tabletypesize{\scriptsize}

\tablecaption{Galaxy Names}

\renewcommand{\arraystretch}{.6}
\tablehead{
\colhead{}&
\multicolumn{3}{c}{RA (J2000)}&
\multicolumn{3}{c}{DEC (J2000)}&
\colhead{}&
\colhead{}&
\colhead{}\\
\colhead{ES Name}&
\colhead{h}&
\colhead{m}&
\colhead{s}&
\colhead{$^o$}&
\colhead{$^{\prime}$}&
\colhead{$^{\prime\prime}$}&
\colhead{Other Name\tablenotemark{a}}&
\colhead{SDSS Name}&
\colhead{Morphological Type\tablenotemark{a}}}
\startdata
HIPEQ0014$-$00  &     00 & 14 & 31.87 & $-$00 & 44& 15.0 &                    UGC00139 &  SDSS J001431.87$-$004415.0 &	 SAB(s)c\\
HIPEQ0027$-$01a  &     00 & 27 & 49.46 & $-$01 & 11& 60.0 &                    UGC00272 &  SDSS J002749.46$-$011160.0 &  SAB(s)d \\
HIPEQ0033$-$01  &     00 & 33 & 21.96 & $-$01 & 07& 18.8 &                    UGC00328 &  SDSS J003321.96$-$010718.8 &	 SB(rs)dm\\
HIPEQ0043$-$00  &      00 & 43 & 27.77 & $-$00 & 07& 30.4 &                    NGC0237 &  SDSS J004327.77$-$000730.4 &	 SAB(rs)cd \\
HIPEQ0051$-$00  &       00 & 51 & 59.59 & $-$00 & 29& 11.8 &                    ARK018 &  SDSS J005159.59$-$002911.8 &	 Sb\\
HIPEQ0058+00 & 00 & 58 & 48.82 & +00 & 35 & 12.1 &                  IC1607/UGC00611 &  SDSS J005848.82+003512.1 &	  Sb \\
HIPEQ0107+01  &  01 & 07 & 46.30 & +01 & 03 & 49.0 &                       UGC00695 &  SDSS J010746.30+010349.0 &	 Sc\\
HIPEQ0119+00  &  01 & 19 & 58.78 & +00 & 43 & 18.5 &                           LSBC\_F827$-$05 &  SDSS J011958.78+004318.5 &  Sd \\
HIPEQ0120$-$00  &  01 & 20 & 06.58 & $-$00 & 12 & 19.1 &                       UGC00866 &  SDSS J012006.58$-$001219.1 &	  Sdm \\
HIPEQ0122+00  &  01 & 22 & 09.10 & +00 & 56 & 44.9 &                        NGC0493 &  SDSS J012209.10+005644.9 &	  SAB(s)cd \\
\enddata

\tablenotetext{a}{Name and morphological type information obtained from NED.}
\label{names}

\end{deluxetable}

\begin{deluxetable}{lrrrrrrrrr}
\tablewidth{0pt}
\tablecolumns{10} 
\tabletypesize{\scriptsize}
\tablecaption{Petrosian Photometry of HI Selected Sources}
\renewcommand{\arraystretch}{.6}
\tablehead{
\colhead{ES Name}&
\colhead{RA (J2000)\tablenotemark{c}}&
\colhead{DEC (J2000)\tablenotemark{c}}&
\colhead{$u$}&
\colhead{$g$}&
\colhead{$r$}&
\colhead{$i$}&
\colhead{$z$}&
\colhead{PetroR50 ($^{\prime\prime}$)}&
\colhead{PetroR90 ($^{\prime\prime}$)}}
\startdata
   HIPEQ0014$-$00\tablenotemark{a} &    3.63280 &    -0.7375 & $14.88\pm{0.04}$ &  $13.92\pm{0.01}$ &  $13.47\pm{0.02}$ &  $13.28\pm{0.01}$ &  $13.06\pm{0.03}$ &   $21.4\pm{0.4}$ &   $59.0\pm{0.8}$\\ 
  HIPEQ0027$-$01a\tablenotemark{a} &    6.95610 &    -1.2000 & $15.65\pm{0.04}$ &  $14.75\pm{0.01}$ &  $14.32\pm{0.02}$ &  $14.16\pm{0.01}$ &  $14.10\pm{0.04}$ &   $21.4\pm{0.4}$ &   $43.6\pm{0.8}$\\ 
   HIPEQ0033$-$01\tablenotemark{b} &    8.34150 &    -1.1219 & $16.06\pm{0.06}$ &  $15.15\pm{0.01}$ &  $14.78\pm{0.02}$ &  $14.61\pm{0.01}$ &  $14.49\pm{0.05}$ &   $20.2\pm{0.4}$ &   $45.9\pm{1.6}$\\ 
   HIPEQ0043$-$00\tablenotemark{a} &   10.86570 &    -0.1251 & $14.50\pm{0.03}$ &  $13.37\pm{0.01}$ &  $12.82\pm{0.02}$ &  $12.55\pm{0.01}$ &  $12.29\pm{0.03}$ &   $13.9\pm{0.4}$ &   $36.4\pm{0.4}$\\ 
   HIPEQ0051$-$00\tablenotemark{a} &   12.99830 &    -0.4866 & $16.08\pm{0.03}$ &  $15.08\pm{0.01}$ &  $14.56\pm{0.02}$ &  $14.34\pm{0.01}$ &  $14.17\pm{0.03}$ &    $6.3\pm{0.4}$ &   $17.4\pm{0.4}$\\ 
   HIPEQ0058+00\tablenotemark{a} &   14.70340 &     0.5867 &   $15.14\pm{0.03}$ &  $14.06\pm{0.01}$ &  $13.53\pm{0.02}$ &  $13.29\pm{0.01}$ &  $13.11\pm{0.03}$ &   $10.7\pm{0.4}$ &   $23.0\pm{0.4}$\\ 
   HIPEQ0107+01\tablenotemark{b} &   16.94290 &     1.0636 &   $15.95\pm{0.04}$ &  $15.07\pm{0.01}$ &  $14.74\pm{0.02}$ &  $14.61\pm{0.01}$ &  $14.55\pm{0.04}$  & $10.7\pm{0.4}$ &   $26.5\pm{0.8}$\\ 
   HIPEQ0119+00\tablenotemark{b} &   19.99490 &     0.7218 &   $18.02\pm{0.11}$ &  $17.17\pm{0.02}$ &  $16.80\pm{0.03}$ &  $16.66\pm{0.03}$ &  $16.50\pm{0.10}$  &  $7.1\pm{0.4}$ &   $15.4\pm{2.0}$\\ 
   HIPEQ0120$-$00\tablenotemark{a} &   20.02740 &    -0.2053 & $16.27\pm{0.04}$ &  $15.33\pm{0.01}$ &  $14.88\pm{0.02}$ &  $14.70\pm{0.01}$ &  $14.53\pm{0.04}$ &   $18.2\pm{0.4}$ &   $35.6\pm{0.8}$\\ 
   HIPEQ0122+00\tablenotemark{a} &   20.53790 &     0.9458 &   $13.68\pm{0.03}$ &  $12.80\pm{0.01}$ &  $12.38\pm{0.02}$ &  $12.11\pm{0.01}$ &  $11.87\pm{0.03}$  & $53.5\pm{0.4}$ &  $105.7\pm{0.8}$\\ 
\enddata
\tablenotetext{a}{Photometry was derived using an elliptical aperture.}
\tablenotetext{b}{Photometry was derived using a circular aperture.}
\tablenotetext{c}{RA and DEC given in decimal degrees.}
\label{table:paper}
\end{deluxetable}


\begin{deluxetable}{lrrrrrrrrrrrrrrrr}
\tablewidth{0pt}
\tablecolumns{17} 
\tabletypesize{\scriptsize}
\tablecaption{Corrections to Petrosian Photometry}
\renewcommand{\arraystretch}{.6}
\tablehead{
\colhead{}&
\multicolumn{5}{c}{Milky Way Extinction\tablenotemark{a}}&
\multicolumn{5}{c}{Internal Extinction\tablenotemark{b}}&
\multicolumn{5}{c}{K-correction\tablenotemark{c}}&
\colhead{Edge}\\
\colhead{ES Name}&
\colhead{$u$}&
\colhead{$g$}&
\colhead{$r$}&
\colhead{$i$}&
\colhead{$z$}&
\colhead{$u$}&
\colhead{$g$}&
\colhead{$r$}&
\colhead{$i$}&
\colhead{$z$}&
\colhead{$u$}&
\colhead{$g$}&
\colhead{$r$}&
\colhead{$i$}&
\colhead{$z$}&
\colhead{Corr.}}
\startdata
 HIPEQ0014$-$00  &       1.09  &       0.79  &       0.53  &       0.46  &       0.29  &       0.85  &       0.62  &       0.45  &       0.34  &       0.24 &       0.02 &      -0.01  &      -0.02  &      -0.01  &      -0.02   &  0.00\\ 
  HIPEQ0027$-$01a  &       0.92  &       0.65  &       0.44  &       0.37  &       0.23  &       0.78  &       0.57  &       0.42  &       0.32  &       0.22 &       0.02 &      -0.01  &      -0.02  &      -0.01  &      -0.02   &  0.00\\ 
   HIPEQ0033$-$01  &       0.39  &       0.29  &       0.20  &       0.17  &       0.10  &       0.28  &       0.20  &       0.15  &       0.11  &       0.08 &      0.00 &      0.00  &       0.00  &      0.00  &      -0.01   &  0.00\\ 
   HIPEQ0043$-$00  &       0.58  &       0.42  &       0.29  &       0.24  &       0.16  &       0.48  &       0.35  &       0.25  &       0.19  &       0.14 &       0.02 &       0.01  &      -0.01  &      0.00  &      -0.01   &  0.00\\ 
   HIPEQ0051$-$00  &       0.65  &       0.47  &       0.32  &       0.29  &       0.18  &       0.45  &       0.33  &       0.24  &       0.18  &       0.13 &       0.00 &      0.00  &       0.00  &      0.00  &      0.00   &  0.00\\ 
   HIPEQ0058$+$00  &       0.21  &       0.16  &       0.09  &       0.10  &       0.05  &       0.02  &       0.01  &       0.01  &       0.01  &       0.00 &       0.02 &       0.02  &       0.01  &       0.00  &      -0.01   &  0.00\\ 
   HIPEQ0107$+$01  &       0.14  &       0.11  &       0.05  &       0.08  &       0.04  &       0.00  &       0.00  &       0.00  &       0.00  &       0.00 &      -0.01 &       0.00  &      0.00  &      0.00  &      0.00   &  0.00\\ 
   HIPEQ0119$+$00  &       0.15  &       0.12  &       0.08  &       0.06  &       0.04  &       0.02  &       0.01  &       0.01  &       0.01  &       0.00 &      -0.01 &      -0.01  &       0.01  &      -0.01  &      -0.01   &  0.00\\ 
   HIPEQ0120$-$00  &       0.57  &       0.42  &       0.29  &       0.26  &       0.17  &       0.39  &       0.28  &       0.21  &       0.16  &       0.11 &      0.00 &      0.00  &       0.00  &      0.00  &      -0.01   &  0.01\\ 
   HIPEQ0122$+$00  &       1.51  &       1.10  &       0.78  &       0.63  &       0.42  &       1.35  &       0.99  &       0.72  &       0.54  &       0.39 &      -0.01 &      -0.01  &      -0.01  &      -0.01  &      -0.02   &  0.00\\ 
\enddata
\tablenotetext{a}{Computed using values in Schlegel, Finkbeiner, \& Davis (1998).}
\tablenotetext{b}{Based on the method of Tully et al. (1998).}
\tablenotetext{c}{Computed using {\tt kcorrect\_v3.2} from Blanton et al. (2003b).}
\label{table:photcor}
\end{deluxetable}

\begin{deluxetable}{lrrrrrrrrrr}
\tablewidth{0pt}
\tablecolumns{11} 
\tabletypesize{\tiny}

\tablecaption{Derived Quantities}

\renewcommand{\arraystretch}{.6}
\tablehead{
\colhead{}&
\colhead{Distance}&
\colhead{Rotational}&
\colhead{Log(Stellar Mass)}&
\multicolumn{5}{c}{Absolute Magnitude}&
\colhead{R50 (kpc)}&
\colhead{R90 (kpc)}\\
\cline{5-9}\\
\colhead{ES Name}&
\colhead{(Mpc)}&
\colhead{Velocity (km\ s$^{-1}$)\tablenotemark{a}}&
\colhead{($\rm{M}_\odot$)}&
\colhead{$u$}&
\colhead{$g$}&
\colhead{$r$}&
\colhead{$i$}&
\colhead{$z$}&
\colhead{$r$-band}&
\colhead{$r$-band}}
\startdata
   HIPEQ0014$-$00  &     56.5$\pm$7.3 &    159.6$\pm$14.0  &     10.0$\pm$0.6  & -19.97$\pm$0.30 & -20.62$\pm$0.27  & -20.82$\pm$0.39 & -20.94$\pm$0.16 & -20.99$\pm$0.30 &   5.85 &  16.15\\ 
  HIPEQ0027$-$01a  &     55.4$\pm$7.2 &    119.1$\pm$13.6  &      9.6$\pm$0.6  & -18.98$\pm$0.30 & -19.62$\pm$0.27  & -19.84$\pm$0.36 & -19.93$\pm$0.20 & -19.85$\pm$0.30 &   5.74 &  11.70\\ 
   HIPEQ0033$-$01  &     27.9$\pm$3.6 &     \nodata\nodata  &      8.8$\pm$0.6  & -16.56$\pm$0.32 & -17.37$\pm$0.28  & -17.64$\pm$0.43 & -17.79$\pm$0.13 & -17.84$\pm$0.32 &   2.73 &   6.21\\ 
   HIPEQ0043$-$00  &     59.5$\pm$7.7 &    191.6$\pm$17.0  &     10.4$\pm$0.6  & -19.95$\pm$0.30 & -20.92$\pm$0.27  & -21.34$\pm$0.49 & -21.56$\pm$0.06 & -21.73$\pm$0.30 &   4.00 &  10.50\\ 
   HIPEQ0051$-$00  &     22.7$\pm$3.0 &     99.1$\pm$14.6  &      8.8$\pm$0.6  & -16.35$\pm$0.30 & -17.18$\pm$0.27  & -17.54$\pm$0.45 & -17.72$\pm$0.10 & -17.79$\pm$0.30 &   0.70 &   1.92\\ 
   HIPEQ0058+00  &     77.5$\pm$10.1 &     \nodata\nodata  &     10.3$\pm$0.6  & -19.51$\pm$0.30 & -20.54$\pm$0.27  & -21.00$\pm$0.52 & -21.25$\pm$0.04 & -21.38$\pm$0.30 &   4.02 &   8.63\\ 
   HIPEQ0107+01  &      8.5$\pm$1.1 &     \nodata\nodata  &      7.7$\pm$0.6  & -13.83$\pm$0.31 & -14.68$\pm$0.27  & -14.95$\pm$0.43 & -15.10$\pm$0.14 & -15.13$\pm$0.31 &   0.44 &   1.09\\ 
   HIPEQ0119+00  &     62.1$\pm$8.1 &     \nodata\nodata  &      8.6$\pm$0.6  & -16.10$\pm$0.37 & -16.91$\pm$0.29  & -17.24$\pm$0.43 & -17.37$\pm$0.17 & -17.50$\pm$0.37 &   2.14 &   4.65\\ 
   HIPEQ0120$-$00  &     24.0$\pm$3.1 &     67.8$\pm$13.3  &      8.7$\pm$0.6  & -16.20$\pm$0.31 & -16.99$\pm$0.27  & -17.31$\pm$0.42 & -17.46$\pm$0.14 & -17.54$\pm$0.31 &   2.12 &   4.14\\ 
   HIPEQ0122+00  &     33.0$\pm$4.3 &    134.2$\pm$12.9  &      9.9$\pm$0.6  & -20.42$\pm$0.30 & -20.89$\pm$0.27  & -20.98$\pm$0.39 & -21.10$\pm$0.17 & -21.14$\pm$0.30 &   8.54 &  16.89\\ 
\enddata
\tablenotetext{a}{Rotational velocities are derived from $W_{20}$ values.  Galaxies without proper inclination corrections have been omitted.}
\label{table:derive}
\end{deluxetable}

\clearpage

\end{landscape}

\end{document}